\documentclass[11pt]{article}
\usepackage{amsmath,epsf,amssymb,latexsym,enumerate,cite,shadow,array,color}
\usepackage{slashed}
\usepackage{graphicx,wasysym}

\DeclareFontFamily{U}{rsf}{} \DeclareFontShape{U}{rsf}{m}{n}{
  <5> <6> rsfs5 <7> <8> <9> rsfs7 <10-> rsfs10}{}
\DeclareMathAlphabet\Scr{U}{rsf}{m}{n} \makeatletter
\@addtoreset{equation}{section} \makeatother

\def\be{\begin{equation}}
\def\ee{\end{equation}}
\def\ba{\begin{array}}
\def\ea{\end{array}}

\newcommand{\bea}{\begin{eqnarray}}
\newcommand{\eea}{\end{eqnarray}}

\def\K{K{\"a}hler}
\def\Re{\mathop{\rm Re}\nolimits}
\def\Im{\mathop{\rm Im}\nolimits}
\newcommand{\hc}{{\rm h.c.}}
\newcommand{\ft}[2]{{\textstyle\frac{#1}{#2}}}
\def\rmi{{\rm i}}
\def\rmd{{\rm d}}
\def\rme{{\rm e}}
\newcommand{\edet}{e\,}
\newcommand{\PLa}{} % to reinstall, change the last to {P_L}
\newcommand{\PRa}{} % to reinstall, change the last to {P_R}
\usepackage{ifpdf}
\ifx\pdfoutput\undefined
   \pdffalse
\else
   \pdfoutput=1
   \pdftrue
  \usepackage[pdftex]{hyperref}
\newcommand{\gamfive}{\gamma_*}
\newcommand{\SU}{\mathop{\rm SU}}

\newcommand{\U}{\mathop{\rm {}U}}

\begin{document}

\begin{titlepage}
\begin{flushright}
{\it Dedicated to the memory of Lev Kofman}
\end{flushright}

\vskip 1cm

\begin{center}
{\LARGE \textbf{ Jordan Frame Supergravity and Inflation in NMSSM \\[0pt]
\vskip 0.8cm }}

{\bf {\bf Sergio Ferrara$^{1,2,3}$},
{\bf Renata Kallosh$^4$}, {\bf Andrei Linde$^4$},\\ {\bf Alessio Marrani$^4$}, and  {\bf Antoine Van Proeyen$^5$} \\

\

$^1${\sl Physics Department, Theory Unit, CERN,  CH 1211, Geneva
23,
Switzerland }\\
$^2${\sl INFN - Laboratori Nazionali di Frascati, Via Enrico
Fermi 40, 00044 Frascati, Italy}\\
$^3${\sl Department of Physics and Astronomy,
University of California, Los Angeles, CA USA}\\
$^4${\sl Department of Physics, Stanford University, Stanford, CA
94305 USA}\\
$^5${\sl Instituut voor Theoretische Fysica, Katholieke Universiteit Leuven,\\ Celestijnenlaan 200D, B-3001 Leuven,
Belgium}}
\end{center}
\vskip 0.5 cm

\begin{abstract}
We present a complete explicit  $N=1$, $d=4$ supergravity action in an
arbitrary Jordan frame with non-minimal scalar-curvature coupling of the
form $\Phi(z, \bar z)\, R$.   The action is derived by suitably
gauge-fixing  the superconformal action.  The theory has a modified \K
\, geometry, and it exhibits a significant dependence on the frame
function $\Phi (z, \bar z)$ and its derivatives over scalars, in the
bosonic as well as in the fermionic part of the action. Under certain
simple conditions, the scalar kinetic terms in the Jordan frame have a
canonical form.

We consider an embedding of the Next-to-Minimal Supersymmetric Standard
Model (NMSSM) gauge theory into supergravity, clarifying  the Higgs
inflation model recently proposed by  Einhorn and Jones.  We find that
the conditions for canonical kinetic terms are satisfied for the NMSSM
scalars in the Jordan frame,  which leads to a simple action. However,
we find that the gauge singlet field experiences a strong tachyonic
instability during inflation in this model. Thus, a modification of the
model is required to support the Higgs-type inflation.

\end{abstract}

\vspace{24pt}
\end{titlepage}

\tableofcontents

\newpage

\section{Introduction}

Supersymmetry imposes certain restrictions on the non-supersymmetric
models of particle physics and cosmology. A well known example of such
restrictions is the fact that the supersymmetric version of the Standard
Model (SM) of particle physics requires at least two Higgs superfields.
Meanwhile, for cosmology Einstein equations have to be solved, therefore
the supersymmetry embedding of the Higgs model inflation requires local
supersymmetry, \textit{i.e.} supergravity. Thus, one can try to see how
the potential discovery of supersymmetry may affect various models of
inflation, derived in the past in the context of general relativity
coupled to scalar fields without supersymmetry. It would be interesting
to find general restrictions, as well as to study particular models.

Here we are  motivated by a particular issue in cosmology, the so-called
$\xi \phi ^{2}R$ coupling, which attracted a lot of attention starting
from the early days of inflation \cite{Futamase:1987ua}. Recently, it
became also quite important in the context of SM inflation \cite{Sha-1}.

Until now, the $N=1$, $d=4$ supergravity action in an arbitrary Jordan
frame described by the frame function $\Phi(z,{\bar z})$, with arbitrary
K{\"a}hler potential $\mathcal{K}(z, {\bar z})$,  holomorphic superpotential
$W(z)$ and holomorphic function $f_{ab}(z)$, was not known.  Here we
will derive this action, which is the first goal of this paper. This
will be achieved by starting with the superconformal theory developed in
\cite{Kallosh:2000ve}, and by gauge-fixing all extra symmetries in order
to get a general  supergravity action in Jordan  frame.

Our results generalize the formulation of $N=1$ supergravity in Jordan
frame for the particular case in which the K{\"a}hler potential
$\mathcal{K}$ and the frame function are related by $\mathcal{K}(z,
{\bar z})= -3 \log (-\ft13 \Phi(z, {\bar z}))$. The corresponding action
in Jordan frame was derived in components in
\cite{Cremmer:1978hn,CFGVP-1}, and in superspace in
\cite{Girardi:1984eq,Wess:1992cp}. In our treatment, we will also
specify the conditions required for the frame function to make the
kinetic terms of the scalar fields canonical in the Jordan frame.

The non-minimal coupling of scalar fields to curvature is allowed by all
known symmetries of the SM and general relativity. If one tries to
describe the early universe using the particle physics SM coupled to
gravity in the Einstein frame, one finds that: 1) the coupling $\lambda$
of the Higgs field has to be of the order $10^{-13}$; 2) the mass of the
Higgs field has to be of the order $10^{13}$ GeV. These conditions may
be satisfied in a general theory of a scalar field, but not in the
simplest version of the standard model. However, if the $\xi \phi ^{2}R$
coupling is included, \textit{i.e.} if the embedding of the particle
physics SM into the Jordan frame gravity is considered, a satisfactory
description of cosmology for the Higgs mass in the interval between
$126$ and $194$ GeV can be found  \cite{Sha-1}.  This is possible for
very large values of the non-minimal scalar-curvature coupling $\xi \sim
10^{4}$. The model predicts the cosmological parameters $n_{s}\approx
0.97$, and $r\approx 0.003$, which are consistent with cosmological
observations. Thus, this model provides very interesting predictions,
which will be testable both at LHC and by a Planck satellite.

When this work was in progress,  a very interesting proposal
\cite{Einhorn:2009bh} was made how to generalize the model of
Bezrukov-Shaposhnikov \cite{Sha-1} in presence of supersymmetry. Under
certain assumptions, it was found that slow regime inflation is not
possible within the supergravity embedding of the Minimal Supersymmetric
Standard Model (MSSM), but rather it is possible for the NMSSM (see
\textit{e.g.} \cite{Ellwanger:2009dp} for a recent review of NMSSM).

In the present paper we will study the supergravity embedding of the
NMSSM and look for a consistent cosmological models of the Higgs-type
inflation.

Firstly, we will derive the complete $N=1$ action in the general Jordan
frame, where it is very simple and has interesting features. This will
help to clarify the meaning of the large non-minimal $\xi \phi ^{2}R$
coupling in the context of supergravity. In particular, the origin of
the canonical kinetic terms of all scalars of the NMSSM in the Jordan
frame is explained, whereas in the Einstein frame scalar kinetic terms
are generally very complicated.

Secondly, we will study the theory as a function of all three chiral
multiplets, namely two Higgs doublets and a singlet,  and analyze
various directions in the space of scalar fields. In particular, in
\cite{Einhorn:2009bh} it was shown that a slow-roll inflationary regime
is possible in NMSSM when the Higgs fields move in the  $D$-flat
direction of the two Higgs doublets $H_u$ and $H_v$, assuming that the
gauge singlet $S$ is small. However, it was not clear whether this last
assumption is justified, {\it i.e.} whether $S = 0$ corresponds to a
minimum of the potential with respect to the field $S$ when inflation
takes place in the $D$-flat direction of the two doublet Higgs fields.
We will show that, unfortunately, the potential of the field $S$ has a
sharp maximum near $S = 0$ in this regime. This means that the
inflationary regime studied in \cite{Einhorn:2009bh} is unstable, and a
search for more general models is required to find a supersymmetric
version of the Higgs-type inflation.

The paper is organized as follows. In Sec. \ref{ss:N1Jordan} we present
the complete explicit  $N=1$, $d=4$ supergravity action in an arbitrary
Jordan frame with non-minimal scalar-curvature coupling of the form
$\Phi(z, \bar z)R$. This includes the bosonic as well as fermionic
action. In the special case in which the frame function $\Phi
(z,\bar{z})$ is related to the K{\"a}hler potential by the relation
$\mathcal{K}(z,{\bar z})= -3 \log (-\ft13 \Phi(z,{\bar z}))$, the action
reduces to the one derived in \cite{Cremmer:1978hn,CFGVP-1}. In the case
$\Phi=-3$, the action becomes the well known action of $N=1$
supergravity in the Einstein frame.

Sec. \ref{ss:BosN1Frames} is devoted to a detailed discussion of the
bosonic part of the supergravity action, which is especially important
for cosmology. In particular, sufficient conditions for the kinetic
terms of scalars to be canonical are specified.

Sec. \ref{ss:sugraNMSSM} starts with a short description of the
Higgs-type inflation with non-minimal scalar-curvature coupling. Then,
we proceed with an attempt to generalize this model to the
supersymmetric case. For this purpose, we study the embedding of the
NMSSM into supergravity, focussing on the Einhorn-Jones cosmological
model \cite{Einhorn:2009bh}. We study this model in the Jordan as well
as in the Einstein frame. The dependence of the potential on the singlet
gauge field $S$, as well as at large values of the Higgs fields in a
$D$-flat direction of the two Higgs doublets, is explicitly computed. We
find that this potential has a maximum for small values of $S$ near the
inflationary trajectory. The resulting instability disallows the
inflationary regime in the model of \cite{Einhorn:2009bh}, unless some
way of stabilizing the field $S$ is found. Sec. \ref{ss:derivN1Jordean}
provides a detailed derivation of the Jordan frame supergravity action
presented in Sec. \ref{ss:N1Jordan}, by gauge-fixing the extra
symmetries of the superconformal action. Finally, the Appendix contains
a discussion of the cosmological behavior of the angle $\beta$ between
the two components of the Higgs field.

\section{Complete $N=1$ Supergravity Action in a Jordan Frame}
 \label{ss:N1Jordan}

The  $N=1$, $d=4$ supergravity action in a Jordan frame with arbitrary
scalar-curvature coupling is uniquely defined by the  frame function
$\Phi(z, \bar z)$, K{\"a}hler potential $\mathcal{K}(z,\bar z)$, holomorphic
superpotential $W(z)$, holomorphic kinetic gauge matrix $f_{AB}(z)$ and
momentum map\footnote{This is also equivalently named Killing potential,
and it encodes the Yang-Mills transformations of the scalars (it may
include Fayet-Iliopoulos terms, as well).} $P_A$. It is given
by\footnote{A derivation of this action, as well as a detailed notation,
is given in Sec. \ref{ss:derivN1Jordean}.} ($e\equiv \sqrt{-g}$)
\begin{eqnarray}
 e^{-1}\mathcal{L}&=&-\ft16\Phi \left[ R(e)-\bar \psi_\mu R^\mu\right] -\ft16
(\partial _\mu\Phi )(\bar \psi \cdot \gamma \psi ^\mu) +\nonumber\\
&&+\mathcal{L}_0+\mathcal{L}_{1/2}+\mathcal{L}_1-V+\mathcal{L}_m+\mathcal{L}_{\mathrm{mix}}+\mathcal{L}_d+\mathcal{L}_{4\mathrm{f}}\,,
 \label{Poincact}
\end{eqnarray}
where the curvature $R(e)$ uses the torsionless connection $\omega _\mu {}^{ab}(e)$,
and the gravitino kinetic term is defined using
\begin{equation}
R^\mu \equiv \gamma ^{\mu \rho \sigma }\left( \partial _{\rho }+\ft14\omega _{\rho }{}^{ab}(e)\gamma _{ab} -\ft32
\rmi\mathcal{A}_\rho \gamfive\right)\psi _\sigma \,.
 \label{defRmuPoinc}
\end{equation}
Here $\mathcal{A}_\mu $ is the part of the auxiliary vector field
containing only bosons, namely:
\begin{equation}
   \mathcal{A}_\mu =\ft16\rmi\left(
\partial _\mu z^\alpha \partial _\alpha \mathcal{K}
-\partial _\mu \bar z^{\bar \alpha }\partial _{\bar \alpha }\mathcal{K}\right)
-\ft13 A_\mu {}^AP_A\,,  \label{Amu}
\end{equation}
where $A_\mu {}^A$ is the Yang-Mills gauge field.

The kinetic terms of spin $0,\frac{1}{2},1$ fields  in (\ref{Poincact})
are respectively given by:
\begin{eqnarray}
 \mathcal{L}_0 & = & -\frac1{4\Phi }(\partial _\mu \Phi )(\partial ^\mu \Phi )
  +\frac13g_{\alpha \bar\beta }\Phi (\hat\partial _\mu z^\alpha )\,(\hat\partial ^\mu\bar z^{\bar\beta })
  \,, \label{Turin-5} \\
 \mathcal{L}_{1/2} & = &-\ft12\tilde g_{\alpha \bar\beta }
   \bar \chi ^{\bar\beta } \PRa \slashed{D} \chi^\alpha
+\ft12\Phi  \bar\chi^\alpha\PLa\gamma ^\mu \chi ^{\bar\beta }\hat{\partial }_\mu z^\gamma
\left[-\ft13 g_{\gamma \bar\beta }L_\alpha+\ft14L_{\alpha\gamma }
L _{\bar\beta }-\ft14L_\alpha L _{\gamma\bar\beta }\right]+\hc\,,
  \label{Turin-1}\\
  \mathcal{L}_1&=&(\Re f_{AB})\left[ -\ft14 F_{\mu \nu }^AF^{\mu \nu
\,B}-\ft12 \bar\lambda ^A {\slashed{D}}\lambda ^B
\right]\nonumber\\ && +\ft 14\,\rmi\left[(\Im f_{AB})\,
 F_{\mu \nu }^A \tilde F^{\mu \nu \,B}+(\hat{\partial}_\mu\Im f_{AB})\,
\bar\lambda ^A \gamfive\gamma ^\mu \lambda^B\right]\,. \label{Turin-6}
\end{eqnarray}
The covariant derivatives of scalars and fermions are defined as
follows:
\begin{eqnarray}
 \hat\partial _\mu z^\alpha &\equiv &\partial _\mu z^\alpha -A_\mu ^A
k_A{}^\alpha\,,\label{Turin-2}\\
\PLa  D_\mu \chi^\alpha  &\equiv &\PLa\left( \partial _\mu+\ft14\omega _\mu {}^{ab}(e)\gamma _{ab}+\ft32\rmi\mathcal{A}_\mu  \right) \chi^\alpha -A_\mu ^A
  \frac{\partial k_A{}^\alpha (z)}{\partial
  z^\beta }\PLa\chi
  ^\beta + \Gamma ^\alpha _{\beta \gamma }
\PLa\chi^\gamma \hat\partial _\mu z^\beta \,,\nonumber\\
D_\mu \lambda ^A &\equiv &\left( \partial _\mu +\ft14 \omega
_\mu {}^{ab}(e)\gamma _{ab} -\ft32\rmi \mathcal{A}_\mu \gamfive\right)
\lambda ^A-A_\mu ^C\lambda ^B f_{BC}{}^A \,.
 \label{covderPoinc}
\end{eqnarray}
The theory has a modified K{\"a}hler geometry. In particular, as given by
(\ref{Turin-1}), the kinetic term of fermions depends on the metric $
  \tilde g_{\alpha \bar\beta }\equiv -\ft13\Phi g_{\alpha \bar\beta }+\ft14\Phi L _\alpha L_{\bar\beta }
 $, where $g_{\alpha \bar\beta }$ is the K{\"a}hler metric and
$
  L_\alpha \equiv \partial _\alpha\ln \left( -\Phi \right) $, $L_{\bar\alpha} \equiv  \partial _{\bar\alpha}\ln \left(-\Phi \right)
=\overline{L_\alpha }$ (see (\ref{deftildeg}) and (\ref{Lalpha}) further
below). Concerning the kinetic terms of scalars, see Sec.
\ref{BosN1-Jordan} . The potential reads
\begin{equation}
  V=\ft19\Phi ^2\left[ \rme^{\mathcal{K}}\left(-3W \overline{W} +\nabla _\alpha Wg^{\alpha \bar\beta }\nabla _{\bar\beta }\overline{W}\right)
 +\ft12(\Re f)^{-1\,AB} P_AP_B\right]  \,.
 \label{VPoincare}
\end{equation}
The fermion mass terms are given by
\begin{eqnarray}
\mathcal{L}_m  &=&\ft12m_{3/2}\bar \psi _\mu P_R
\gamma ^{\mu \nu }\psi _\nu-\ft12m_{\alpha \beta }\bar
\chi^\alpha \PLa\chi^\beta
-m_{\alpha A}\bar \chi^\alpha \PLa\lambda ^A-\ft12m_{AB}\bar\lambda ^AP_L\lambda^B +\hc\,,\nonumber\\
m_{3/2}&=& \left( -\ft13\Phi \right) ^{3/2} \rme^{\mathcal{K}/2}W\,,\nonumber\\
m_{\alpha \beta } &=&\left( -\ft13\Phi \right) ^{3/2} \rme^{\mathcal{K}/2}\left[ \nabla _\beta \nabla _\alpha W
+2L_{(\alpha }\nabla _{\beta )}W\right] \,,\nonumber\\
 m_{AB}&=&-\ft12(-\ft13\Phi)^{1/2}\rme^{\mathcal{K}/2}f_{AB\,\alpha } g^{\alpha \bar\beta }\overline\nabla_{\bar\beta }\overline{W}\,,\nonumber\\
   m_{\alpha A} &=&
  -\ft13\rmi\sqrt{2}\Phi\left[  \left( \partial  _\alpha +\ft12L_\alpha \right)P_A-\ft1{4} f_{AB\,\alpha }(\Re f)^{-1\,BC} P_C\right]\,.
\label{Lmferm}
\end{eqnarray}
The remaining terms read
\begin{eqnarray}
\mathcal{L}_{\mathrm{mix}}&=& \bar  \psi  \cdot \gamma P_L\left[ -\frac16 \rmi\,\Phi \,P_A
\lambda ^A+\frac{1}{\sqrt{2}}\chi ^\alpha \rme^{\mathcal{K}/2}
 \left( \partial _\alpha +(\partial _\alpha K)\right) (-\ft13\Phi )^{3/2} W\right]+\hc\,,\nonumber\\
\mathcal{L}_d &=& \ft18(\Re f_{AB})\bar \psi _\mu \gamma ^{ab}\left( F_{ab}^A+ \widehat F_{ab }^A \right)
\gamma ^\mu \lambda ^B  \nonumber\\
&&+\frac{1}{\sqrt{2}}\left\{\bar \psi_\mu\PLa\gamma ^\nu \gamma^\mu\chi ^\alpha
\left[ (-\ft13\Phi ) g_{\alpha \bar\beta }\hat{\partial }_\nu\bar z^{\bar\beta }+\ft14 L _\alpha
\partial _\nu \Phi
   \right] \right. \nonumber\\
&&\left.\ - \ft1{4}f_{A B\,\alpha } \bar \chi^\alpha \PLa\gamma ^{ab}
\widehat
F_{ab}^{-A } \lambda ^B-\ft13\Phi L_\alpha \bar\chi^\alpha \PLa\gamma ^{\mu \nu} D_\mu \psi_\nu+\hc\right\}\,,
 \label{parts}
\end{eqnarray}
where
 $\widehat{F}_{ab}{}^A\equiv e_a{}^\mu e_b{}^\nu\left( 2\partial _{[\mu }A_{\nu
]}^A +g f_{BC}{}^AA^B_\mu A^C_\nu+\bar \psi _{[\mu } \gamma_{\nu
]}\lambda ^A\right)$ (see (\ref{confgaugemult})). The explicit
expression for the 4-fermion terms $\mathcal{L}_{4\mathrm{f}}$ will
be presented in Sec. \ref{ss:derivN1Jordean}. Remarkably, also
$\mathcal{L}_{4\mathrm{f}}$ contains a significant dependence on the
frame function $\Phi$ and its derivatives.

\section{Bosonic Action of $N=1$ Supergravity in Einstein and Jordan Frames}
 \label{ss:BosN1Frames}

\subsection{The Einstein frame}\label{BosN1-Einstein}

By setting $\Phi =-3$ in (\ref{Poincact}), the general $N=1$ action in a
Jordan frame reduces to the well known action of $N=1$ supergravity in
the Einstein frame \cite{Cremmer:1978hn,CFGVP-1}.

It is here worth recalling some basic facts about the structure of the
bosonic sector of $N=1$, $d=4$ supergravity. In $M_P=1$ units, the
action of $N=1$ supergravity coupled to chiral and vector matter
multiplets is usually given in  the Einstein frame, where the curvature
$R$ appears in the action only through the Einstein-Hilbert term
${\frac{1}{2}} \sqrt{-g_E} \,R(g_{E})$, where $g_{E}^{\mu \nu }$ is the
Einstein frame space-time metric. The theory is defined by a real K{\"a}hler
function $\mathcal{K}(z,{\bar z})$, by an holomorphic superpotential
$W(z)$ and by an holomorphic matrix $f_{AB}(z)$ defining the action of
the vector multiplets \cite{CFGVP-1}.  A particular feature of the
theory is the K{\"a}hler geometry of the complex scalar fields.

The purely bosonic Lagrangian density reads
\begin{equation}
\mathcal{L}_{E}^{bos}= \mathcal{L}_{E}^{grav} +\mathcal{L}_{E}^{scalar}+  \mathcal{L}_{E}^{vec}\,, \label{Einstein-1}
\end{equation}
where
\begin{eqnarray}
&&\mathcal{L}_{E}^{grav}+\mathcal{L}_{E}^{scalar}=\sqrt{-g_{E}}\left[ {\ft{1}{2}}R(g_{E})-g_{\alpha\bar\beta}\hat{\partial}
_\mu z^\alpha \hat{\partial}_\nu\bar z^{\bar\beta }g_{E}^{\mu \nu }-V_E\right]\,,
\label{Einstein-2}\\
&&\mathcal{L}_{E}^{vec}=\sqrt{-g_{E}}\left[  -\ft14 \left( \Re  f_{AB}\, \right) F_{\mu\nu}^AF^{\mu \nu \,B}
+\ft14 \rmi\left(  \Im f_{AB}\,\right)  F_{\mu\nu}^A\tilde F^{\mu\nu \, B}\right]\,.
\label{vec}
\end{eqnarray}
Note that  the contractions of spacetime indices and the definition of
the dual field strength  are performed using the  Einstein  frame metric
$g_{E}^{\mu\nu }$. The strictly positive-definite metric
$g_{\alpha\bar\beta }(z,\bar z)$ of the non-linear sigma model of
scalars $z^\alpha,\bar z^{\bar\beta }$ is given by the second derivative
of the real K{\"a}hler potential
\begin{equation}
g_{\alpha\bar\beta }(z,\bar z)\equiv {\frac{\partial }{\partial z^\alpha}}{\frac{%
\partial }{\partial \bar z^{\bar\beta }}}\mathcal{K}(z,\bar z)>0\,,
\label{Kibark}
\end{equation}
and $\hat{\partial}_\mu z^\alpha$ is the Yang-Mills gauge covariant
derivative of a scalar field, defined by (\ref{Turin-2}).

Concerning the potential $V_{E}$, the $F$-term potential $V_{E}^{F}$
depends on $\mathcal{K}$ and $W$. On the other hand, the $D$-term
potential $V_{E}^{D}$ depends on  the values of the auxiliary
$D$-fields, obtained by solving the corresponding equations of motion:
\begin{equation}
V_{E}=V_{E}^F +V_{E}^D = \rme^{\mathcal{K}}\left( -3W\overline{W}
+\nabla _\alpha Wg^{\alpha \bar\beta }\nabla _{\bar\beta }\overline{W}
\right) +  \ft12(\Re
f)^{-1\,AB}P_AP_B\,. \label{VE}
\end{equation}
Notice that (\ref{VE}) yields that
\begin{equation}
V_{J}=\frac{\Phi ^{2}}{9}V_{E}\,,  \label{Turin-3}
\end{equation}
where the potential in Jordan frame $V_{J}\equiv V$ is given by
(\ref{VPoincare}). $\nabla ^\alpha W$ denotes the K{\"a}hler-covariant
derivative of the superpotential. The $D$-term potential can be
presented also in the form $V_{E}^D= \ft12 \left( \Re  f^{AB} \right)
D^A D^B$, where $D^A$ is the value of the auxiliary field of the vector
multiplets. This is the standard form of the purely bosonic part of the
$N=1$, $d=4$ supergravity action in the Einstein frame. Of course, such
a bosonic action can be made supersymmetric by adding suitable fermionic
terms (see \textit{e.g.} \cite{Cremmer:1978hn,CFGVP-1}).

\subsection{The Jordan frame}\label{BosN1-Jordan}

In order to find the action in an arbitrary Jordan frame, one can
perform a change of variables from the Einstein to the Jordan frame.
Only the metric and the fermions have to be rescaled, the scalars and
the vector fields do not change. The metric in a Jordan frame  is
related to the metric in the Einstein frame as follows (subscripts
``$E$'' and ''$J$'' respectively stand for Einstein and Jordan
frames throughout)
\begin{equation}
   g_{J}^{\mu\nu}=\Omega^2 g_{E}^{\mu\nu}\,,\qquad \Omega^2 = - \ft13 \Phi (z,{\bar z})>0\,.
\label{Turin-3-bis}
\end{equation}
Within our treatment, we will consider the  scale factor $\Omega^2$ as
an arbitrary real function of the complex scalar fields $(z,{\bar z})$.
Its positivity, through (\ref{Turin-3-bis}), correspondingly constrains
$\Phi (z,{\bar z})$. Since the new action in a Jordan frame is related
to the standard one in the Einstein frame by a change of variables, it
is supersymmetric, as the original one.

Instead of performing the above change of variables by ``brute force'',
in Sec. \ref{ss:derivN1Jordean} we use as a starting point an $N=1$,
$d=4$ superconformal theory \cite{Kallosh:2000ve} with local
$\SU(2,2|1)$ symmetry. Such a superconformal theory has a set of local
symmetries which includes all $N=1$ supergravity symmetries and, in
addition, a set of extra local symmetries: local dilatation,  $\U(1)$
symmetry and special supersymmetry. The superconformal theory has no
dimensionful  parameters.

In \cite{Kallosh:2000ve}  the local dilatation,  $\U(1)$ symmetry and
special supersymmetry were gauge fixed in a way that allowed to
reproduce the standard $N=1$, $d=4$ supergravity action in the Einstein
frame. In fact, the purely bosonic action of $N=1$ supergravity in a
Jordan frame is already suggested by  Eq. (C.5) of
\cite{Kallosh:2000ve}. The complete $N=1$ supergravity action in $d=4$
in a generic Jordan frame has been presented in Sec. \ref{ss:N1Jordan},
and it is thoroughly derived in Sec. \ref{ss:derivN1Jordean} through a
suitable gauge-fixing of superconformal supergravity theory
\cite{Kallosh:2000ve}. This is a symmetry-inspired approach, alternative
to the ``brute force'' computation based on the change of variables
(\ref{Turin-3-bis}). Here we will just present the results for the
purely bosonic part of the supergravity action in a Jordan frame, which
is the most relevant one for cosmology.

As mentioned above, the locally supersymmetric action is defined by the
choice of four independent  functions: a real K{\"a}hler potential
$\mathcal{K}(z,{\bar z})$, an holomorphic superpotential $W(z)$ and an
holomorphic matrix $f_{AB} (z)$, determining the kinetic vector matrix.
This suffices to define the $N=1$, $d=4$ supergravity in the Einstein
frame. When dealing with a Jordan frame, an additional fourth function,
namely the real frame function $\Phi (z,{\bar z})$, has to be specified.
Thus, the purely bosonic part of the $N=1$, $d=4$ supergravity in a
generic Jordan frame reads
\begin{eqnarray}
\mathcal{L}_{J}^{bos} &=& \sqrt{-{g}_{J}}\left[ -\frac{1}{6}\Phi {R}({g}_J)+\left( \frac{1}{3}\Phi g_{\alpha\bar\beta }-\frac{\Phi
_\alpha \Phi _{\bar\beta }}{\Phi }\right) \hat{\partial}_\mu z^\alpha \hat{\partial}^\mu \bar z^{\bar\beta }
-\frac{\left( \Phi _\alpha \hat{\partial}z^\alpha -\Phi _{\bar\beta }\hat{\partial}\bar z^{\bar\beta
}\right)^2}{4\Phi }
-\frac{\Phi^2}{9}V_E + \mathcal{L}_{1}^{bos}\right]\,.\nonumber\\
&&  \label{Turin-4}
\end{eqnarray}
Here $V_E$ is the Einstein frame potential defined in (\ref{VE}) $\frac{
\Phi^2}{9}V_{E}=V_{J}\equiv V$ is the Jordan frame potential given by
(\ref{VPoincare}), and
\begin{equation}
 \Phi_\alpha \equiv {\frac{\partial }{\partial z^\alpha }}\Phi(z, \bar z) \,, \qquad  \Phi_{\bar\beta } \equiv {\partial\over \partial{\bar z}^{\bar\beta }}  \Phi(z, \bar z)=\overline{\Phi _\beta }\,.
\end{equation}
Notice that (\ref{Turin-4}) is implied by (\ref{Poincact}), (\ref
{Turin-5}), and (\ref{Turin-2}), observing that $\partial _\mu \Phi
=\hat{
\partial}_\mu \Phi $ because in general $\Phi $ is gauge-invariant.
Furthermore,
$\mathcal{L}_{1,J}^{bos}=\mathcal{L}_{1,E}^{bos}=\mathcal{L} _{1}^{bos}$
is conformal invariant (and therefore frame independent), and it is
given by the purely bosonic part of (\ref{Turin-6}), or equivalently by
($\left( -g_{E}\right) ^{-1/2}$ times) (\ref{vec}):
\begin{equation}
\mathcal{L}_{1,J}^{bos}=\mathcal{L}_{1,E}^{bos}=\mathcal{L}_{1}^{bos}=-\ft14\left(  \Re f_{AB} \right)   F_{\mu\nu}^AF^{\mu\nu \,B}+\ft14\rmi\left(\Im
  f_{AB}\right) \,F_{\mu\nu}^A\tilde F^{\mu\nu \,B}\,.
\label{vec1}
\end{equation}
In the Jordan frame, the contractions of space-time indices and the
definition of the dual field strength are performed using the Jordan
frame metric $g_{J}^{\mu\nu}$ given by (\ref{Turin-3-bis}).

It should be remarked that (\ref{Turin-4}) yields that the geometry of
the non-linear sigma model of scalars is of a modified K{\"a}hler type:
indeed, due to the term proportional to $\left( \Phi _\alpha
\hat{\partial} z^\alpha-\Phi _{ \bar\beta }\hat{\partial}\bar
z^{\bar\beta }\right)^2$, the metric is not Hermitian, i.e. there are
terms of the form $\rmd z\rmd z$ and complex conjugate; furthermore, the
metric term of $\rmd z\rmd \bar z$ is not of the K{\"a}hler type.

As a consequence of the previous treatment and computations, by setting
$ \Phi =-3$ in (\ref{Turin-4}) the purely bosonic part of the $N=1,$
$d=4$ supergravity action in the Einstein frame
\cite{Cremmer:1978hn,CFGVP-1} is recovered. With the choice $\Phi
=-3\rme^{-\ft13 \mathcal{K}(z,{\bar z})}$, (\ref{Turin-4}) yields to the
purely bosonic action of $N=1$ supergravity in the particular Jordan
frame considered in \cite{Cremmer:1978hn,CFGVP-1}.

\subsection{Canonical kinetic terms for scalars}\label{Canonical-Suff-Conds}

In the Einstein frame, the kinetic term of scalar fields is given by
$g_{\alpha\bar\beta } \hat{\partial} _\mu z^\alpha \hat{\partial}^\mu
\bar z^{\bar\beta } $, where $g_{\alpha\bar\beta }(z,\bar z)$ is given
by (\ref{Kibark}). Thus, canonical kinetic terms are possible for the
following choice of a K{\"a}hler potential:
\begin{equation}
\mathcal{K}(z,{\bar z})=\delta_{\alpha\bar\beta }z^{\alpha} \bar z^{\bar\beta } +f(z) +\bar f({\bar z})\,,
\label{Turin-7}
\end{equation}
where $f(z)$ is a holomorphic function (associated to the considered
K{\"a}hler gauge). A $1$-modulus example of the canonical K{\"a}hler potential
(\ref{Turin-7}) is provided by the shift-symmetric function
$\mathcal{K}(z,{\bar z})=-\ft12(z-\bar z)^2$, often used in cosmology.

As pointed out above, an early version of  $N=1$, $d=4$ supergravity
theory in a (particular) Jordan, as well as in the Einstein, frame was
derived in \cite{Cremmer:1978hn,CFGVP-1} on the basis of the
superconformal calculus, with $\Phi $ and $\mathcal{K}$ related as
follows:
\begin{equation}
\mathcal{K}(\Phi(z,{\bar z}))= -3\log (-\ft13\Phi(z,{\bar z}))\,,
\label{K}
\end{equation}
and $\Omega^2$ given by (\ref{Turin-3-bis}).

Within such a framework, the following simpler form of $\mathcal{L}
_{J}^{bos} $ given by (\ref{Turin-4}) is obtained:
\begin{equation}
\mathcal{L}_{J}^{bos,\mathcal{K}(\Phi )}=\sqrt{-g_{J}}\left[ -{\textstyle
\frac{1}{6}}\Phi R(g_{J})-\Phi _{\alpha\bar\beta }\hat{\partial}
_\mu z^\alpha\hat{\partial}^\mu \bar z^{\bar\beta }+\Phi
\mathcal{A}_\mu^2-\frac{\Phi^2}{9}V_{E}+\mathcal{L}_{1}^{bos,
\mathcal{K}(\Phi )}\right] \,.  \label{Turin-11}
\end{equation}
The kinetic term for the scalar action is partly determined by the value
$\mathcal{A}_\mu $ of the bosonic part of the auxiliary field of
supergravity, entering in the action $\mathcal{L}_{J}^{bos}$
(\ref{Turin-4}) as $\Phi \mathcal{A}_\mu \mathcal{A}^\mu$. In the case
of gauge-invariant $\mathcal{K}$, $\mathcal{A}_\mu $ reads\footnote{When
the  K{\"a}hler potential is not gauge-invariant in direction $A$, the
auxiliary pseudovector has an additional contribution depending on a
gauge field, $+\ft16\rmi A_\mu^A(r_A-\bar r_A)\,$, where $r_A$ is the
holomorphic part of the transformation of the K{\"a}hler potential under
gauge symmetry, $\delta \mathcal{K}(z,{\bar z})= \theta^A\left[ r_A(z)+
\bar r_A({\bar z})\right]$, see \cite{Kallosh:2000ve}. }
\begin{equation}
 \mathcal{A}_\mu=  \frac16\rmi\,
  \left( \hat{\partial} _\mu z^\alpha \partial_\alpha \mathcal{K}
  - \hat{\partial} _\mu \bar z^{\bar\alpha  }\partial_{\bar\alpha }\mathcal{K}\right)= -\frac{\rmi}{2\Phi }\,
  \left( \hat{\partial}_\mu z^\alpha\partial_\alpha\Phi
  - \hat{\partial} _\mu \bar z^{\bar\alpha }\partial_{\bar\alpha }\Phi \right) \,.
 \label{calAmugeneral}
\end{equation}

The purely bosonic action (\ref{Turin-11}) yields to the following
statement: within the relation (\ref{K}) between $\mathcal{K}$ and $\Phi
$, \textit{in order to have canonical kinetic terms in the Jordan frame
it is sufficient}

 \textbf{a}) to choose the frame function $\Phi $ as follows:
\begin{equation}
 \Phi(z,{\bar z}) =-3 + \delta_{\alpha\bar\beta }z^{\alpha} \bar z^{\bar\beta } +J(z) +\bar J({\bar z})\,,
\label{Turin-12}
\end{equation}
where $J(z)$ is holomorphic. Note that, through (\ref{K}),  (\ref
{Turin-12}) implies $\mathcal{K}$ to read:
\begin{equation}
\mathcal{K}(z,{\bar z})=-3\log \left[ 1-\ft{1}{3}\delta _{\alpha\bar{
\alpha}}z^\alpha\bar z^{\bar{\alpha}}-\ft{1}{3}J(z)-\ft{1}{3}\bar{J}(
{\bar z})\,\right] \,;
\end{equation}

\textbf{b}) to consider only (scalar) configurations for which the
contribution from the bosonic part of the auxiliary vector field
vanishes:
\begin{equation}
\mathcal{A}_\mu =0\,.  \label{Turin-13}
\end{equation}

The embedding of the NMSSM into supergravity along the lines suggested
in \cite{Einhorn:2009bh} requires only the knowledge of the simple case
in which the relation (\ref{K}) between $\mathcal{K}$ and $\Phi $ holds.
Moreover, concerning the canonicity of the kinetic terms of scalars, in
the treatment below we will see that condition \textbf{a}) is always
satisfied, and condition \textbf{b}) given by (\ref{Turin-13}) is
satisfied during the cosmological evolution, when the system under
consideration depends on three real fields: $h_1, h_2, s$. Thus, apart
from the frame function $\Phi $ given by (\ref{Turin-12}), the action of
the NMSSM embedded in supergravity in Jordan frame (\ref{K}) along the
lines of \cite{Einhorn:2009bh} has canonical kinetic scalar terms and a
potential $\frac{\Phi^2}{9}V_{E}$ (see Secs. \ref{EJ-Embedding},
\ref{Steste} for details). In particular, when only the Higgs field $h$
is non-vanishing in the D-flat direction, the Jordan frame supergravity
potential is extremely simple and is given by  ${\lambda^2 \over 4}
h^4$, see Eq. (\ref{Ste-10}).

\section{Supergravity embedding of the NMSSM and Cosmology}
 \label{ss:sugraNMSSM}
\subsection{Classical approximation of the Higgs-type inflation with non-minimal $\protect\xi$-coupling}

The essential reason for the new version of the SM inflation
\cite{Sha-1} to work successfully  is the following. The SM potential
with canonical kinetic term for the Higgs field $h$ is coupled to a
gravitational field in a suitable Jordan frame. In other words, the
Lagrangian density to start with reads:
\begin{equation}
\mathcal{L}_{J}=\sqrt{-g_{J}}\left[ {\frac{M^2+\xi h^2}2
}R\left( g_{J}\right) -\ft12\partial _\mu h\partial _\nu hg_{J}^{\mu\nu}-{\frac{\lambda
}{4}}(h^2-v^2)^2\right] \,. \label{Trn-1}
\end{equation}
At present, $h = v \sim 10^{-16} M_P$, and $M_P^2 = M^2 + \xi v^2$.
Therefore $M\approx M_P$ for $\sqrt \xi <10^{16}$. In the subsequent
investigation we will consider $\xi < 10^6$. In this case $M = M_P$ with
a very good accuracy. In our paper we will use the system of units where
$M = M_P = 1$.

In general, the cosmological predictions have to be compared with the
observations in the Einstein frame, related to the Jordan one through
the conformal rescaling
 (\ref{Turin-3-bis}), with
\begin{equation}
\Phi =-3 (1+\xi h^2)\,,\qquad \Omega^2={1+\xi h^2}\,.
\end{equation}
By switching to the Einstein frame, (\ref{Trn-1}) yields to
\begin{equation}
\mathcal{L}_E=\sqrt{-g_{E}}\left( \ft12 R(g_{E})-\ft12\partial _\mu \psi \partial _\nu\psi\, g_{E}^{\mu\nu}-U(\psi )\right)\,,
\end{equation}
where $\psi $ is a canonically normalized scalar in the Einstein frame,
defined by
\begin{equation}
\rmd\psi \equiv  \rmd h \sqrt {\Omega^2 + 6 \xi^2 h^2\over \Omega^4}\,.
\end{equation}
where
\begin{equation}
U(\psi )={\frac{\lambda }{4}}\left( {\frac{h^2(\psi
)-v^2}{{ {1 + \xi
h(\psi )^2}}}}\right)^2\,. \label{U}
\end{equation}

The relation between the field $h$ and the canonically normalized field
$\psi$ looks very different in {\it three} different ranges of $h$.  At
$h\ll \frac{1}{\xi }$ one has  $\psi \approx h$.  In the interval
${\frac{1}{{\xi }}} \ll h\ll {\frac{1}{\sqrt{\xi }}}$, the relation
between $h$ and $\psi$ is more complicated: $\psi \approx \sqrt{3\over
2}\, \xi h^2$. Finally, for $h \gg {\frac{1}{\sqrt{\xi }}}$ (or,
equivalently, $\psi \gg 1$) one has $h\sim {\frac{1}{\sqrt{\xi }}}\,
e^{\frac{\psi }{\sqrt{6}}} $. In this regime,  the potential in the
Einstein frame is very flat, which leads to inflation:
\begin{equation}
U(\psi )  _{\psi \rightarrow \infty } \;  \Rightarrow \;
{
\frac{\lambda }{4\xi^2}}\left( 1+\rme^{-{\frac{2\psi }{
\sqrt{6}}}}\right)^{-2}\,.  \label{Trn-2}
\end{equation}
As one can see from (\ref{Trn-2}), the constant ($\psi $-independent)
term in the potential $U\left(\psi \right)$ is $\frac{\lambda}{4\xi^2}$,
so nothing would work without the non-minimal scalar curvature coupling
proportional to $\xi$.

The Hubble constant during inflation in this model is $H \approx
{\sqrt{\lambda\over 3}}\, {1\over 2\xi}$. For the non-supersymmetric
standard model, $\lambda = O(1)$, so one could worry that this energy
scale is dangerously close to the possible unitarity bound $\Lambda \sim
1/\xi$ discussed in
\cite{Burgess:2009ea,Barbon:2009ya,Hertzberg:2010dc}. One should note,
however, that most of the arguments suggesting the existence of this
bound are based on the investigation of the theory in the small field
approximation $\psi \approx h$, where one can use an expansion $\psi  =
h (1 + \xi^2 h^2+...)$. This approximation is valid only for $h\ll
{\frac{1}{{\xi }}}$, which is parametrically far from the inflationary
regime at $h \gg {\frac{1}{\sqrt{\xi }}}$. We are going to return to
this issue in a forthcoming publication; see also a discussion in
\cite{Lerner:2009na}, and especially in \cite{Lee:2010hj}, where it was
noticed that in NMSSM one may consider the regime with $\lambda \ll 1$,
where the concerns about the unitarity bound do not seem to appear.

It is worth noting that potentials exponentially rapidly approaching a
constant positive value have been proposed in one of the first models of
chaotic inflation in supergravity \cite{Goncharov:1983mw}, but at that
time models of this type were lacking a compelling motivation.
Therefore, it is very tempting to use the intuitively appealing and
simple model discussed above as a starting point, in order to analyze
the Einhorn-Jones approach \cite{Einhorn:2009bh} to embed NMSSM into
$N=1$, $d=4$ supergravity, and its relevance for the issue of inflation.

\subsection{Embedding of the NMSSM into Supergravity and the Einhorn-Jones cosmological inflationary model}
\label{EJ-Embedding}

The Higgs field sector of NMSSM has one gauge singlet and two gauge
doublet chiral superfields, namely \cite{Ellwanger:2009dp}:
\begin{equation}
z^{\alpha}=\left\{ S,H_{1},H_{2}\right\} \,,
\end{equation}
with
\begin{eqnarray}
S&=&s\rme^{\rmi\alpha }\,, \nonumber\\
H_u &=&\left(
\begin{array}{c}
H^+_{u} \\
H^0_{u}
\end{array}
\right)  \, , \qquad H_d=\left(
\begin{array}{c}
H^0_{d} \\
H_{d}^-
\end{array}
\right)\,.
\end{eqnarray}
As in \cite{Einhorn:2009bh}, the frame function is chosen as follows:
\begin{equation}
   \Phi (z,{\bar z})=  -3+ (S \bar S + H_u H_u^\dagger + H_d H_d^\dagger) +\ft32 \chi (H_u\cdot  H_d + \hc)\,,
\label{Ste-1}
\end{equation}
where
\begin{equation}
H_{u}\cdot H_{d}\equiv -H_{u}^{0}H_{d}^{0}+H_{u}^{+}H_{d}^{-}.
\end{equation}
Note that (\ref{Ste-1}) is of the form (\ref{Turin-12}), with
$J=\frac{3}2 \chi H_{u}\cdot H_{d}$. In this framework, the K{\"a}hler\
potential is related to $\Phi $ through (\ref{K}), and the
superpotential is chosen to be
\begin{equation}
W=-\lambda SH_{u}\cdot H_{d}+{\rho\over 3} S^{3}.
\end{equation}
Thus, the action of such an implementation of NMSSM depends on five
chiral superfields. Through explicit computations, we checked that such
an action admits a consistent truncation in which the charged
superfields, namely $ H_{u}^{+}$ and $H_{d}^{-}$, are absent. Therefore,
below we deal with a simplified action of NMSSM, containing only three
superfields: $S$, $ H_{u}^{0}$ and $H_{d}^{0}$, such that:
\begin{equation}
H_{1}=\left(
\begin{array}{c}
0 \\
H_{u}^{0}
\end{array}
\right) \,,\qquad H_2=\left(
\begin{array}{c}
H_{d}^{0} \\
0
\end{array}
\right) \,.
\end{equation}
Within this truncation, the frame function and the superpotential
respectively read:
\begin{eqnarray}
{\textstyle}\Phi (z,{\bar z}) &=&-3+{\textstyle}\left(
|S|^2+|H_{u}^{0}|^2+|H_{d}^{0}|^2\right)
-\ft32\chi
(H_{u}^{0}H_{d}^{0}+\overline{H_{u}^{0}}\overline{H_{d}^{0}})\,;
\label{frame} \\
W &=&\lambda SH_{u}^{0}H_{d}^{0}+{\rho\over 3} S^{3}\,.  \label{W-NMSSM}
\end{eqnarray}

Thus, by recalling Eqs. (\ref{Turin-3}), (\ref{Turin-11}) and
(\ref{Turin-12} ), and by disregarding
$\mathcal{L}_{1}^{bos,\mathcal{K}(\Phi )}$ in (\ref {Turin-11}), one
obtains the following Jordan frame supergravity scalar-gravity action
for this implementation of NMSSM:
\begin{equation}
{\frac{\mathcal{L}_{J}^{NMSSM}}{\sqrt{-g_{J}}}}=\ft12
R(g_{J})+\ft16\left[\delta _{\alpha\bar\beta }z^\alpha\bar z^{\bar\beta }+\ft32\chi (H_{u}^{0}H_{d}^{0}+
\overline{H_{u}^{0}}\overline{H_{d}^{0}})\right]R-\delta _{\alpha\bar\beta
}\hat{\partial}_\mu z^\alpha\hat{\partial}^\mu \bar z^{\overline{
\beta }}-\Phi \mathcal{A}_\mu^2-V_{J}\,,  \label{Ste-3}
\end{equation}
where $\mathcal{A}_\mu $ is given by (\ref{calAmugeneral}).

Remarkably, the scalar-curvature coupling exhibited by (\ref{Ste-3})
breaks the discrete $\Bbb{Z}_{3}$ symmetry of the theory due to the
chosen cubic superpotential (\ref{W-NMSSM}) of NMSSM. Such a symmetry
may generate domain walls after the spontaneous breaking of a symmetric
phase in the early universe. In such a case, unacceptably large
anisotropies of CMB may be generated. This is a well known domain wall
problem of NMSSM (see \textit{ e.g.} \cite{Ellwanger:2009dp}). The
scalar-curvature coupling in (\ref{frame} ) and in (\ref{Ste-3}) breaks
the discrete $\Bbb{Z}_{3}$ symmetry. This may help to remove the
eventual domain wall problem. Thus, it is challenging and interesting to
formulate a consistent cosmology within this framework.

As usual, $V_{J}=V_{J}^{F}+V_{J}^{D}$. In the present framework,
$V_{J}^{F}$ has a zero, second and fourth power of the $S$ field:
\begin{equation}
V_{J}^{F}=\lambda^2|H_{u}^{0}|^2|H_{d}^{0}|^2+\lambda \rho (\bar{S}
^2H_{u}^{0}H_{d}^{0}+c.c.)-{\frac{2\lambda
^2|S|^2|H_A^{0}|^2(\chi
(H_{u}^{0}H_{d}^{0}+c.c.)-2)}{
4+3\chi^2|H_A^{0}|^2-2\chi (H_{u}^{0}H_{d}^{0}+h.c.) }}
+\rho^2|S|^{4}\,.  \label{Ste-4}
\end{equation}
On the other hand, $V_{J}^{D}$ reads
\begin{equation}
V_{J}^{D}={\frac{g^{^{\prime }2}}{8}}(|H_{u}^{0}|^2-|H_{d}^{0}|^2)^2+{
\frac{g^2}{8}}((H_{u})^{\dagger }\vec{\tau}H_{u}+(H_{d})^{\dagger }\vec{
\tau}H_{d})^2\,,  \label{Ste-5}
\end{equation}
where $\vec{\tau}$ is the $3$-vector of Pauli $\sigma $-matrices.

In \cite{Einhorn:2009bh} this model was described at the vanishing value
of the gauge singlet field $S$. In order to analyze the theory
consistently, in the present treatment we keep the full dependence on
$S$.

\subsection{Cosmology in the Jordan frame} \label{Steste}

We start by checking that the $CP$-invariant solution found in
\cite{Einhorn:2009bh}, in which $S$, $H_{u}^{0}$ and $H_{d}^{0}$ are
real, corresponds to a(n at least local) minimum of $V_{J}$ itself. In
order to do so, \textit{a priori} we assume that these three fields are
complex, namely ( $s,h_{1},h_2\in \Bbb{R}^{+}$, $\alpha ,\alpha
_{1},\alpha _2\in \left[ 0,2\pi \right) $):
\begin{equation}
S=s\rme^{\rmi\alpha }\,,\qquad H_{u}^{0}=h_{1}\rme^{
\rmi\alpha _{1}}\,,\qquad
H_{d}^{0}=h_2\rme^{\rmi\alpha _2}\,.
\end{equation}
By computing (\ref{Ste-3}), it follows that the scalar-gravity action
depends only on the combination angles $\gamma \equiv \alpha _{1}+\alpha
_2 $ and $\delta \equiv 2\alpha -\alpha _{1}-\alpha _2$. More precisely,
the dependence on $\delta $ enters via $\lambda \rho \cos \delta $ and
the dependence on $\gamma $ is via $\chi \cos \gamma $. In order to
study $CP$ -invariant solution(s) with $\alpha =\alpha _{1}=\alpha
_2=0$, one has to analyze the minima of the potential $V_{J}$, also
taking into account the $R$ -dependent terms in (\ref{Ste-3}) (notice
that $V_{J}^{D}$ does not depend on any phase).

Firstly, we notice that Eqs. (\ref{Ste-4}), (\ref{Ste-5}) and the
definition of $\delta $ yield that the dependence on $\delta $ enters
only in one term in the potential, namely:
\begin{equation}
V_{J}(\delta )=2\lambda \rho |S|^2|H_{u}^{0}||H_{d}^{0}|\cos
\delta \,.
\end{equation}
This potential has a minimum at $\delta =0$, under the condition that $
\lambda \rho $ is negative: $\lambda \rho =-|\lambda \rho |$.

Secondly, in order to deal correctly with the dependence on $\gamma $,
one can look at the expected minimum of the potential at $S=0$ \cite
{Einhorn:2009bh}. (\ref{Ste-4}) implies that the Jordan frame potential
at $ S=0$ is very simple:
\begin{equation}
\left. V_{J}^{F}\right| _{S=0}=\lambda
^2|H_{u}^{0}|^2|H_{d}^{0}|^2\,.
\end{equation}
At $S=0$ the dependence on $\gamma $ enters only through the frame
function $ \Phi $ given by Eq. (\ref{frame}). By switching to the
Einstein frame, and recalling the relation (\ref{Turin-3}), one obtains:
\begin{equation}
(V_{E}(\gamma ))_{S=0}={\frac{9\lambda^2|H_{u}^{0}|^2|H_{d}^{0}|^2}{
\Phi^2}}={\frac{\lambda^2|H_{u}^{0}|^2|H_{d}^{0}|^2}{\left[ 1-{
\textstyle\frac{1}{3}}\left( |H_{u}^{0}|^2+|H_{d}^{0}|^2\right)
+\chi |H_{u}^{0}||H_{d}^{0}|\cos \gamma \right]^2}}\,.
\end{equation}
Since during inflation $1-\ft13(|H_A^{0}|^2)>0$ \cite {Einhorn:2009bh},
it can be checked that during inflation $\gamma =0$ is a minimum of
$V_{J}$, under condition that $\chi >0$.

Thus, the $CP$-invariant solution with three real fields is confirmed to
be a minimum in the directions of angles $\delta $ and $\gamma $ during
inflation. Therefore we can take
\begin{equation}
S=s\,,\qquad H_{u}^{0}=h_{1}\,,\qquad H_{d}^{0}=h_2,
\label{Ste-7}
\end{equation}
provided that the coupling constants of the model under consideration
satisfy
\begin{equation}
\lambda \rho <0\,,\qquad \chi >0\,.
\end{equation}

Notice that (\ref{frame}) and (\ref{Ste-7}) yield that the kinetic
scalar terms in the Jordan frame are canonical, since both sufficient
conditions ( \ref{Turin-12}) and (\ref{Turin-13}) are satisfied (in
particular, $\mathcal{ A}_\mu =0$ on scalar configurations
(\ref{Ste-7})):
\begin{equation}
(\mathcal{L}_{J}^{NMSSM})_{kinetic}=-\sqrt{-g_{J}}\left[ (\partial
_\mu s)^2+(\partial _\mu h_{1})^2+(\partial _\mu h_2)^2\right] \,.
\end{equation}
It is now convenient to switch to the standard mixing of the Higgs
fields, defined as:
\begin{equation}
h_{1}\equiv h\cos \beta \,,\qquad h_2\equiv h\sin \beta \,,
\label{ansatz}
\end{equation}
which leaves us with two real fields, $h$ and $\beta $, instead of
$h_{1}$ and $h_2$.

Through Eq. (\ref{Ste-5}), the $D$-flat direction, defined by
\begin{equation}
V_{J}^{D}=0,  \label{Ste-8}
\end{equation}
requires that
\begin{equation}
\sin (2\beta )=1;~~~~h_{1}^2=h_2^2=h^2/2.  \label{Ste-9}
\end{equation}
Thus, along the $D$-flat direction, the curvature term of (\ref{Ste-3})
simplifies to:
\begin{equation}
(\mathcal{L}_{J}^{NMSSM})_{curv}={\frac{\sqrt{-g_{J}}}2}\left[ 1-{
\textstyle\frac{1}{3}}\left( s^2+h^2\right)
+\ft12\chi h^2\right] R(g_{J})\,.
\end{equation}

On the other hand, along the $D$-flat direction
(\ref{Ste-8})-(\ref{Ste-9}) the $F$-term potential reads
\begin{equation}
V_J^F={\frac{\lambda^2}{4}}h^{4}- |\lambda \rho|s^2h^2-{\frac{
2\lambda^2s^2h^2(\chi h^2-2)}{ 4+3\chi ^2h^2-2\chi h^2
}}+\rho^2s^{4}\,.\label{Ste-10}
\end{equation}

In \cite{Einhorn:2009bh} the inflationary regime driven by the Higgs
within NMSSM was shown to take place for
\begin{equation}
\chi h^2\gg 1
\gg h^2\,,\qquad s\approx 0\,,\qquad \beta ={\frac{\pi }{4}}
\,,  \label{infl}
\end{equation}
in Planck units $M_P^2=1$. For small $s$, (\ref{Ste-10}) can be
simplified as follows:
\begin{equation}
V_{J}^{F}\sim {\frac{\lambda^2}{4}}h^{4}-\left(|\lambda \rho |+{\frac{2\lambda^2}{3\chi }}\right) s^2h^2\,.
\end{equation}
The effective mass of the $s$ field is negative, but one actually has to
take into account an effective contribution from the curvature-scalar
coupling.  This latter provides a positive contribution, however, it
does not remove the tachyonic instability of the system in the $s$
direction. Indeed, for small $s$, the complete expression of the
effective potential is
 \begin{eqnarray}
\tilde{V}_{J}^{F}\sim {\frac{\lambda
^2}{4}}h^{4}-\left(|\lambda \rho |+{\frac{\lambda^2}{3\chi
}}\right) s^2h^2\,.
\end{eqnarray}
As we will see in the next Sec., the instability in the $s$ direction is
very strong, corresponding to a large tachyonic mass and a slow-roll
parameter $|\eta |\geq 2/3$. As a result, a rapidly developing tachyonic
instability does not allow inflation to occur in the regime studied in
\cite{Einhorn:2009bh}.

Note that in general instead of $\lambda \rho <0$ one could take
$\lambda \rho >0$. Correspondingly, such a choice of coupling constants
would stabilize the real part of the field $S$, but it would lead to an
equally strong instability in the direction of its imaginary part. In
other words, independently of the sign of $\lambda \rho $, the potential
with respect to the complex field $S$ has a saddle point at $S=0$, which
results in the tachyonic instability in one of the two directions.

\subsection{Switching to the Einstein frame}\label{Steste2}
\begin{figure}[t]
\centering
\includegraphics[scale=0.4]{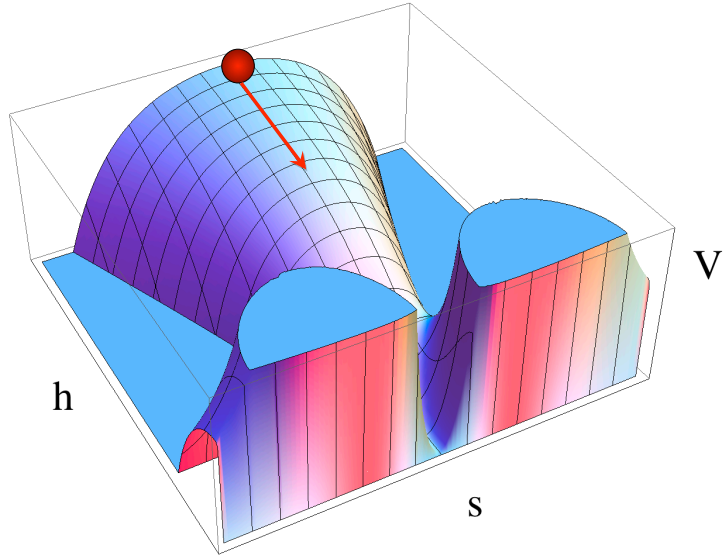}
\caption{The $F$-term potential $V^{F}$ in the Einstein frame.  The inflationary trajectory $s=0$ is unstable.}
\label{fig1}
\end{figure}

In the Einstein frame, (\ref{Turin-3}) and (\ref{Ste-4}) yield that the
$F$-term potential is
\begin{equation}
V_{E}^{F}={\frac{9}{\Phi^2}}V_{J}={\frac{{\frac{\lambda^2}{4}}
h^{4}-|\lambda \rho |s^2h^2-{\frac{2\lambda^2s^2h^2(\chi h^2-2)
}{4+3\chi^2h^2-2\chi h^2 }}+\rho^2s^{4}}{\left[ 1- \ft13\left(
s^2+h^2\right) +\ft12 \chi h^2\right]^2}}\,.
\end{equation}
Let us compute the effective mass of the $s$ field also in the Einstein
frame, where by definition there is no contribution from the curvature
coupling. During the inflationary regime (\ref{infl})
\cite{Einhorn:2009bh}, the leading behavior of the potential is
\begin{equation}
V_{E}^{F}\sim {\frac{\lambda^2}{\chi
^2}}-\left(|\lambda \rho |+{\frac{\lambda^2}{3\chi }}\right)
{\frac{4s^2}{\chi^2h^2}}+O(s^{4})\,.
\end{equation}
The shape of the potential is shown in Fig. \ref{fig1}. The trajectory
with $ s=0$ at large $h$, which was expected to be an inflationary
trajectory in \cite{Einhorn:2009bh}, is unstable. It corresponds to the
top of the ridge for the potential $V_{E}^{F}$, see Fig. \ref{fig1}.

In order to find whether this instability is dangerous, one should
calculate the tachyonic mass of the $s$ field and compare it to the
Hubble constant. This will allow us to check whether the tachyonic
instability develops rapidly, or whether it occurs on a time scale much
smaller than the cosmological time scale $H^{-1}$. An alternative way to
approach this issue is to find the related value of the relevant
slow-roll parameter $\eta $.

To find the effective mass of the $s$ field, attention must be paid to
the non-minimal normalization of the field $S=s\rme^{\rmi\alpha }$. At
constant $\alpha $, the kinetic term of field $S$ is given by
\begin{equation}
g_{S\overline{S}}\partial S\partial \bar{S}={\frac2{\chi
h^2}}\partial S\partial \bar{S}={\frac2{\chi h^2}}(\partial
s)^2\,.
\end{equation}
Thus, in the vicinity of the inflationary trajectory $s\approx 0$
(\ref{infl} ), the Lagrangian density of the field $s$ is
\begin{equation}
\mathcal{L}_{E,s}= -{\frac2{\chi h^2}}(\partial s)^2-{
\frac{\lambda^2}{\chi^2}}+\left(|\lambda \rho |+{\frac{\lambda^2}{
3\chi }}\right) {\frac{4s^2}{\chi^2h^2}} + {\cal O} (s^4)\,.
\end{equation}
In terms of the canonical scalar field $\tilde s={\frac{2s}{\sqrt{\chi
}h}}$, such a Lagrangian at small fields $\tilde s$ is
\begin{equation}
\mathcal{L}_{E,\widetilde{s}}= -{\textstyle\frac{
1}2}(\partial \tilde{s})^2-{\frac{\lambda^2}{\chi^2}}+\left( {
\frac{|\lambda \rho |}{\chi }}+{\frac{\lambda^2}{3\chi
^2}}\right) \tilde{s}^2 + {\cal O} (\tilde s^4) \,,
\end{equation}
resulting in the mass squared of the $\tilde{s}$ field to be
tachyonic:
\begin{equation} m_{\tilde{s}}^2\sim -2\left( {\frac{\lambda^2}{3\chi^2}}+{\frac{|\lambda \rho |}{\chi }}\right) <0\,.
\end{equation}
Taking into account that during inflation $H^{2} = V/3 \approx
{\frac{\lambda^2}{3\chi^2}}$, it thus follows that
\begin{equation}
m_{\tilde{s}}^2\leq -{\frac{2\lambda^2}{3\chi^2}}=-{\frac{2V}{3}}
=-2H^2=R/6\,.
\end{equation}
Interestingly, $m_{\tilde{s}}^2$ resembles the conformal mass
$m^2=-R/6$, but has an opposite sign. Since $|m_{\tilde{s}}^2|>H^2$, the
trajectory $\tilde{s}=0$ is exponentially unstable and unsuitable for
inflation. One can also reach the same conclusion by computing the
relevant slow-roll parameter $\eta $ in the $\tilde{s}$ direction:
\begin{equation}
\eta _{\widetilde{s}}\equiv {\frac{m_{\tilde{s}}^2}{V}}=- {\frac2{3
}}-{\frac{2|\lambda \rho |\chi }{\lambda^2}}
<-{\frac2{3}}\,.
\end{equation}

We did not find any way to solve this problem of the Einhorn-Jones model
\cite{Einhorn:2009bh}.

It should also be clearly stated that there are many other scalar fields
in this model, and the field $s$ is not the only one which may
experience a tachyonic instability. This is supported by the results
obtained in the Appendix, where the dependence of the potential on the
angular variable $ \beta $ is studied. Therein, we find that in certain
cases the post-inflationary cosmological trajectory may experience an
additional tachyonic instability, and deviate from the value $\beta
=\frac{\pi }{4}$ characterizing the $D$-flat direction (\ref{infl}).

We should emphasize, however, that these results are model-dependent. We
believe that the cosmological models based on $N=1$, $d=4$ supergravity
in Jordan frame can be very interesting, and they certainly deserve
further investigation. In the past, a systematic study of such models
was precluded by the absence of the corresponding formalism, which we
presented in a complete form in Sec. \ref{ss:N1Jordan}. In the next
Section we will give a detailed derivation of the complete $N=1$, $d=4$
supergravity action in a generic Jordan frame.

\section{Derivation of the Complete $N=1$ Supergravity Action in a Jordan Frame}
 \label{ss:derivN1Jordean}
Here we use the superconformal action \cite{Kallosh:2000ve} and gauge
fix it to get a complete $N=1$ supergravity action, including fermions,
in an arbitrary Jordan frame. Superconformal invariance means that the
action is invariant under the local symmetries of the superconformal
algebra. This involves, apart from the super-Poincar{\'e} transformations,
local dilatations, a local $U(1)$ $R$-symmetry, local special conformal
transformations, and an extra special supersymmetry, denoted as
$S$-supersymmetry. One first constructs a ``superconformal action'',
\textit{i.e.} an action that is invariant under all symmetries of the
superconformal algebra. Then one gets rid of the extra symmetries by
imposing gauge conditions.

The vierbein $e_\mu^a$ and gravitino $\psi _\mu $ are the gauge fields
of the translations and $Q$-supersymmetry, which belong to the
super-Poincar{\'e} algebra. The gauge field of local Lorentz rotations is
the spin connection $\omega_\mu {}^{ab}$ which is a constrained field,
i.e. it has as usual a value that depends on $e_\mu {}^a$ and $\psi _\mu
$. We will write here the expressions in terms of $\omega_\mu
{}^{ab}(e)$ which is the usual torsionless spin connection of gravity.
Also the gauge fields of special conformal transformations and of
$S$-supersymmetry are such composite fields. In the expressions below,
they have been substituted by their values. On the other hand, the gauge
field of the $U(1)$ $R$-symmetry, $A_\mu $, is an auxiliary field. It is
value will be given below. Finally, the gauge field of dilatations is a
field $b_\mu $, which will later be set to zero by a gauge condition for
the special conformal symmetry.

The superconformal transformations of the vierbein and gravitino are (apart
from general coordinate transformations)
\begin{eqnarray}
\delta e_\mu {}^{a} &=&-\lambda^{a}{}_{b}e_\mu {}^{b}-\lambda _{\mathrm{D}}e_\mu {}^{a}+\ft12\bar{\epsilon}\gamma^{a}\psi _\mu \,, \nonumber\\
\delta \psi _\mu  &=&\left( -\ft14\lambda^{ab}\gamma
_{ab}-\ft12\lambda _{\mathrm{D}}+\ft32
\rmi\lambda _{T}\gamma _{\ast }\right) \psi _\mu +\left( \partial
_\mu +\ft12b_\mu +\ft14\omega _\mu {}^{ab}\gamma _{ab}-\ft32
\rmi A_\mu \gamma _{\ast }\right) \epsilon -\gamma_\mu\eta \,,
  \label{sctransf_e_psi}
\end{eqnarray}
where $\lambda^{ab}$ are the parameters of local Lorentz
transformations, $\lambda _{\mathrm{D}}$ are those of dilatations,
$\lambda _{T}$ are those of the $U(1)$ $R$-symmetry. $\epsilon $ and
$\eta $ are the spinor parameters of $Q$ and $S$-supersymmetry,
respectively.

\subsection{The Superconformal Action}

We first repeat the result for the full superconformal action using the
notation that we will use in this paper. The action contains 3
superconformal-invariant terms
\begin{equation}
\mathcal{L}= [\mathcal{N} ]_D + [\mathcal{W}]_F + \left[f_{AB }\bar\lambda^A P_L \lambda^B\right]_F  \label{totalsconfaction}
\,.
\end{equation}
The first one is defined by a K{\"a}hler potential $\mathcal{N}(X,\bar X)$
for the superconformal fields, the second uses a superpotential
$\mathcal{W}(X)$, and the third involves the chiral kinetic matrix
$f_{AB}(X)$ (where $A$ are the gauge indices), and gauginos $\lambda^A$.
The matrix $P_L=\frac12(1+\gamma _*)$ projects on the left-handed
fermions. The dilatation symmetry implies that $\mathcal{N}$ should be
homogeneous of first order in both $X$ and $\bar X$, $\mathcal{W}$
should be homogeneous of third degree and $f_{AB}(X)$ is of zeroth
order, i.e.
\begin{equation}
X^I\frac{\partial}{\partial X^I}\mathcal{N}=\bar X^{\bar I}\frac{\partial
} {\partial\bar X^{\bar I}}\mathcal{N}=\mathcal{N}\,,\qquad X^I\frac{
\partial}{\partial X^I}\mathcal{W}=3\mathcal{W}\,, \qquad X^I\frac{\partial
}{\partial X^I}f_{AB}=0\,.  \label{homogNWf}
\end{equation}

The superconformal chiral multiplets contain the bosonic fields $X^I$
and fermions $\Omega^I=P_L\Omega^I$. We assume that they transform under
the gauge symmetries depending on Killing vectors $k_A{}^I(X)$
\begin{equation}
\delta X^I= \theta^A k_A{}^I\,,\qquad \delta \PLa\Omega^I = \theta^A \partial_J k_A{}^I \PLa\Omega^J\,.  \label{delgaugechirallocal}
\end{equation}
These Killing vectors should satisfy homogeneity equations due to the
conformal symmetry, and leave $\mathcal{N}$ and $\mathcal{W}$
invariant\footnote{Note that this does not imply that the K{\"a}hler
potential or superpotential of the Einstein theory should be invariant
under the gauge transformations, as we will see below.}. These
statements can be encoded in the following equations
\begin{eqnarray}
& & \partial_{\bar J}k_A{}^I=0\,,\qquad X^J\partial_J k_A{}^I=k_A{}^I\,,
\nonumber\\
& & \mathcal{N}_I k_A{}^I+\mathcal{N}_{\bar I}k_A{}^{\bar I}=0\,,  \nonumber\\
& & \mathcal{P}_A = \ft12\rmi\left(\mathcal{N}_I
k_A{}^I-\mathcal{N}_{\bar I}k_A{}^{\bar I}\right) = \rmi\mathcal{N}_I
k_A{}^I=-\rmi\mathcal{N}_{\bar I}k_A{}^{\bar I}\,,\qquad \partial
_{\bar I}\mathcal{P}_A= \rmi\mathcal{N}_{J\bar I}k_A{}^J\,,  \nonumber\\
&& \mathcal{W}_I k_A{}^I=0\,.  \label{eqnskconform}
\end{eqnarray}
We use here the notation that derivatives on $\mathcal{N}$ and $\mathcal{W}$
are denoted by adding indices, similar to (\ref{Kibark}).

The physical fields of the chiral and gauge multiplets transform as follows
under the superconformal transformations:
\begin{eqnarray}
\delta X^I & = &\left( \lambda _{\mathrm{D}} +\rmi\lambda _{T}
\right) X^I+\frac{1}{\sqrt{2}} \bar\epsilon \PLa\Omega^I \,,  \nonumber\\
\delta \PLa\Omega^I & = &\left( -\ft14\lambda^{ab}\gamma _{ab}+\ft32\lambda _{\mathrm{D}} -\ft12\rmi\lambda _{T}\right) \PLa\Omega^I
+ \frac1{\sqrt{2}} P_L\left(\slashed{\cal D} X^I+F^I\right)\epsilon +\sqrt{2} X^I P_L\eta \,,  \nonumber\\
\delta A^A_\mu &=& -\ft12\bar{\epsilon}\gamma_\mu
\lambda^A\,,  \nonumber\\
\delta\lambda^A &=& \left( -\ft{1}{4}\lambda^{ab}\gamma _{ab}+\ft{3}{2}
\lambda _{\mathrm{D}} +\ft12\rmi\lambda _{T} \gamma _*\right) \lambda^A+
\left[ \ft32\lambda _{\mathrm{D}}+\ft32
\rmi\gamma _*\lambda _{T}\right]\lambda^A + \left[\ft14
\gamma^{ab}\widehat{F}_{ab}{}^A +\ft12\rmi \gamma_*
D^A\right]\epsilon\,,  \nonumber\\
\mathcal{D}_\mu X^I &=& \left( \partial_\mu -b_\mu - \rmi
A_\mu\right) X^I -\frac{1}{\sqrt{2}}\bar \psi _\mu \PLa\Omega^I-A_\mu
^Ak_A{}^I \,, \nonumber\\
\widehat{F}_{ab}{}^A&=&e_a{}^\mu e_b{}^\nu\left( 2\partial_{[\mu }A_{\nu
]}^A +g f_{BC}{}^AA^B_\mu A^C_\nu+\bar \psi _{[\mu } \gamma_{\nu ]}\lambda
^A\right)\,.  \label{confgaugemult}
\end{eqnarray}
After elimination of the auxiliary fields, the terms in
(\ref{totalsconfaction}) mix. The scalars form a K{\"a}hler manifold with
metric, connection and curvature given by
\begin{equation}
  G_{I\bar J}={\cal N}_{I\bar J}\,,\qquad \Gamma^I_{JK}= G^{I\bar L}{\cal N}_{JK\bar L}\,,\qquad
  R_{I\bar K J\bar L}= {\cal N}_{IJ\bar K\bar L}-{\cal N}_{IJ\bar M}G^{M\bar M}{\cal N}_{M\bar K\bar L}\,.
 \label{embeddingKahler}
\end{equation}
The superconformal action\footnote{There is a possible generalization
including a Chern-Simons term, see \cite{DeRydt:2007vg}, which we
neglect here.} can be split in several parts
\begin{eqnarray}
 e^{-1}\mathcal{L}&=&\ft16{\cal N}\left[ -R(e,b) + \bar \psi_\mu R^\mu+ \edet^{-1}
\partial_\mu (\edet \bar \psi \cdot \gamma \psi^\mu)\right]
\nonumber\\
&&\mathcal{L}_0+\mathcal{L}_{1/2}+\mathcal{L}_1-V+\mathcal{L}_m+\mathcal{L}_{\mathrm{mix}}+\mathcal{L}_d+\mathcal{L}_{4\mathrm{f}}\,.
 \label{confact3}
\end{eqnarray}
The leading kinetic terms of the matter multiplets are
\begin{eqnarray}
 \mathcal{L}_0 & = & -G_{I\bar J}D^\mu X^I\,D_\mu \bar X^{\bar J}\,, \nonumber\\
 \mathcal{L}_{1/2} & = & -\ft12G_{I\bar J}\left[
   \bar \Omega^I \PLa \hat{\slashed{D}} \Omega^{\bar J}
  + \ft12\bar \Omega^{\bar J} \PRa \hat{\slashed{D}} \Omega^I \right]\,,
  \nonumber\\
  \mathcal{L}_1&=&(\Re f_{A B})\left[ -\ft14 F_{\mu\nu }^A F^{\mu\nu
\,B } -\ft12 \bar\lambda^A {\slashed{D}}\lambda^B
\right]\nonumber\\ && +\ft 14\,\rmi\left[(\Im f_{A B})\,
 F_{\mu\nu }^A \tilde F^{\mu\nu \,B }+ (D_\mu\Im f_{A B})\,
\bar\lambda^A \gamfive\gamma^\mu \lambda^B\right]\,. \label{confactkin}
\end{eqnarray}
The potential in the conformal form is
\begin{equation}
  V\,=\, V_F+ V_D= G^{I\bar J}{\cal W}_I \overline{{\cal W}}_{\bar J}
 +\ft12(\Re f)^{-1\,A B} {\cal P}_A {\cal P}_B\,.
 \label{confactpot}
\end{equation}
Bilinear fermion terms can be divided in those that give rise to
physical masses, terms relevant for the super-BEH mechanism and terms
with derivative couplings to bosonic fields:
\begin{eqnarray}
\mathcal{L}_m&=& \ft12 {\cal  W}\bar \psi _\mu P_R \gamma^{\mu\nu }\psi
_\nu
 -\ft12\nabla _I{\cal  W}_J\bar \Omega^I\PLa\Omega^J
 +\ft14G^{I\bar J}\overline{{\cal W}}_{\bar J} f_{AB I}\bar\lambda^AP_L\lambda^B\nonumber\\
&&+\sqrt{2}\, \rmi \left[ -\partial_I{\cal P}_A
+\ft1{4}f_{ABI}(\Re f)^{-1\,BC} {\cal P}_C\right]
\bar\lambda^A \PLa\Omega^I
+  \hc\,,\nonumber\\
\mathcal{L}_{\mathrm{mix}}&=&\bar  \psi  \cdot \gamma P_L\left( \frac12 \rmi{\cal P}_A
\lambda^A +\frac{1}{\sqrt{2}}{\cal  W}_I  \Omega^I\right)+\hc\,,\nonumber\\
\mathcal{L}_d &=&  +\frac18(\Re f_{A B})\bar \psi _\mu \gamma^{ab
}\left( F_{ab}^A+ \widehat F_{ab}^A \right)
\gamma^\mu \lambda^B  + \frac{1}{\sqrt{2}}\left\{G_{I\bar J} \bar \psi_\mu \PLa\slashed{D}
X^{\bar J}\,
 \gamma^\mu \Omega^I \right.\nonumber\\ &&\left.\ - \ft1{4}f_{A B\,I} \bar \Omega^I\PLa\gamma^{ab }
\widehat
F_{ab }^A \lambda^B
  -\ft2{3}{\cal N}_I\bar \Omega^I \PLa  \gamma^{\mu\nu} D_\mu \psi_\nu +\hc
\right\}\,.
\label{confact2fermion}
\end{eqnarray}
Finally, the 4-fermion terms are
\begin{eqnarray}
\mathcal{L}_{4\mathrm{f}}&=&\frac{1}{96}{\cal N}\left[ (\bar{\psi}%
^\rho\gamma^\mu\psi^\nu) ( \bar{\psi}_\rho\gamma_\mu\psi_\nu +2 \bar{\psi}%
_\rho\gamma_\nu\psi_\mu) - 4 (\bar{\psi}_\mu \gamma\cdot\psi)(\bar{\psi}^\mu
\gamma\cdot\psi)\right]\nonumber\\
&&+\left\{ - \frac1{4\sqrt{2}}f_{A B\,I}\bar \psi \cdot\gamma  \PLa\Omega
^I \bar\lambda^A P_L\lambda^B
 +\frac18 \nabla _If_{A B\, J }\bar \Omega^I\PLa\Omega^J \bar\lambda^AP_L
 \lambda^B + \hc \right\}\nonumber\\
 &&+\ft{1}{16}\edet^{-1}\varepsilon^{\mu\nu\rho\sigma}\bar
 \psi _\mu \gamma_\nu \psi _\rho\left(\bar \Omega^{\bar J}\PRa\gamma _\sigma \Omega^I
 +\ft12\rmi\Re f_{AB}\bar\lambda^A\gamfive\gamma _\sigma \lambda^B\right)  -\ft12G_{I\bar J}
 \bar \psi _\mu \PRa\Omega^{\bar J}\,\bar \psi _\mu \PLa\Omega^I\nonumber\\
&&  +\ft14R_{I\bar KJ\bar L}\bar \Omega^I \PLa\Omega^J\,\bar \Omega
^{\bar K} \PRa\Omega^{\bar L}-\ft1{16}G^{I\bar J}f_{A B \,I}\bar\lambda^A P_L\lambda^B\bar f_{CD \,\bar J}
\bar\lambda^C P_R\lambda^D \nonumber\\
 &&+\ft1{16}(\Re f)^{-1\,AB}
\left( f_{AC\,I}\bar \Omega^I \PLa - \bar f_{AC\,\bar I}
\bar\Omega^{\bar I }\PRa \right)\lambda^C \left( f_{BD\,J}\bar \Omega^J \PLa
-\bar f_{BD\,\bar J }
\bar\Omega^{\bar J}\PRa\right) \lambda^D
 +{\cal N} (A_\mu^{\rm F})^2  \,.
 \label{confact4fermion}
\end{eqnarray}
This superconformal action contains the bosonic and fermionic parts of
the auxiliary field $A_\mu $, which are
\begin{eqnarray}
\mathcal{A}_\mu&=&\rmi\frac{1}{2\mathcal{N}}\left[ \mathcal{N}_{\bar
I}\,\left( \partial_\mu \bar X^{\bar I}-A_\mu^A k_A{}^{\bar I}\right) -
\mathcal{N}_I\left( \partial_\mu X^I-A_\mu^A k_A{}^I\right) \right]  \nonumber\\
&=& \rmi\frac{1}{2\mathcal{N}}\left[\mathcal{N}_{\bar I}\,{\partial}
_\mu \bar X^{\bar I}-\mathcal{N}_I\,{\partial} _\mu X^I \right]+\frac{1}{%
\mathcal{N}}A_\mu^A \mathcal{P}_A\,,\nonumber\\
A_\mu^{\rm F}&\equiv &\rmi\frac{1}{4{\cal N}}\left[ \sqrt{2}\bar \psi_\mu\left({\cal N}_I
\PLa\Omega^I-{\cal N}_{\bar I} \PRa\Omega^{\bar I}
\right)+G_{I\bar J}\bar \Omega^I\PLa\gamma _\mu \Omega
^{\bar J}  +\frac{3}{2} (\Re f_{A B }) \bar\lambda^A \gamma _\mu \gamfive\lambda^B\right]  \,.  \label{fullAmu}
\end{eqnarray}
$R(e,b)$ is defined with the spin connection $\omega _\mu {}^{ab}(e,b)$,
which is intermediate between the full connection $\omega
_\mu{}^{ab}(e,b,\psi )$ and the torsionless one $\omega _\mu {}^{ab}(e)$:
\begin{eqnarray}
\omega _\mu{}^{ab}(e,b,\psi )&=&\omega _\mu {}^{ab}(e,b)+\ft12\psi _\mu \gamma^{[a}\psi^{b]}+\ft14\bar \psi^a\gamma
_\mu \psi^b \,,\nonumber\\
  \omega _\mu {}^{ab}(e,b)&=& \omega _\mu {}^{ab}(e)+2e_\mu {}^{[a}e^{b]\nu }b_\nu
 \,,\qquad\omega _\mu {}^{ab}(e)= 2 e^{\nu[a} \partial_{[\mu} e_{\nu]}{}^{b]} -
e^{\nu[a}e^{b]\sigma} e_{\mu c} \partial_\nu e_\sigma{}^c\,.
 \label{solomega}
\end{eqnarray}
Fermion terms are extracted from covariant derivatives $D_\mu $, whose
superconformal $U(1)$ connection involves only the bosonic part
$\mathcal{A}_\mu$. Thus, explicitly,
\begin{eqnarray}
D_\mu X^I & = & \partial_\mu X^I-b_\mu X^I-A_\mu
^A k_A {}^I -\rmi \mathcal{A}_\mu
X^I \,, \nonumber\\
\hat{D}_\mu\Omega^I&=&\left(
\partial_\mu -\ft32b_\mu+\ft14 \omega _\mu {}^{ab}(e,b)\gamma _{ab}
 +\ft 12\rmi \mathcal{A}_\mu\right)\PLa\Omega^I-A_\mu^A \partial_Jk _A {}^I\,\PLa\Omega^J
 +\Gamma^I_{JK}\Omega^K D_\mu X^J\,,  \nonumber\\
D_\mu \lambda^A &=&\left( \partial_\mu-\ft32b_\mu +\ft14 \omega
_\mu {}^{ab}(e,b)\gamma _{ab} -\ft 32\rmi \mathcal{A}_\mu \gamfive\right)
\lambda^A  -A_\mu^C\lambda^B f_{BC}{}^A \,.  \label{covderconf}
\end{eqnarray}
We also defined in a similar way
\begin{equation}
R^\mu \equiv \gamma^{\mu \rho \sigma }\left( \partial_{\rho }+\ft12b_\mu  +{\textstyle%
\frac{1}{4 }} \omega _{\rho } {}^{ab}(e,b)\gamma _{ab} -{\textstyle\frac{3}{2}}%
\rmi \mathcal{A}_\rho \gamfive\right)\psi _\sigma \,,
 \label{defRmu}
\end{equation}
while $D_\mu \psi _\nu$ contains also $\psi $-torsion in the derivative.
The action (\ref{confact3}) is invariant under the superconformal
transformations. We now will break those symmetries that are not
required for super-Poincar{\'e} supergravity: special conformal
transformations, dilatations and $S$-supersymmetry.

\subsection{Partial gauge fixing and modified K{\"a}hler geometry}
 \label{ss:partialGaugeFixing}

First, we eliminate the special conformal transformations, by imposing
{\it the special conformal transformations gauge choice}:
\begin{equation}
 b_\mu =0\,.
\label{Kgauge}
\end{equation}
Next, we discuss the gauge choice for dilatations. The {\it dilatational
gauge}, D-gauge, that has been chosen in the past is \cite{Kugo:1982mr}
\begin{equation}
\mathcal{N}=-\frac{3}{\kappa^2}\,.
\label{DgaugeN1matter}
\end{equation}
This brings the Einstein-Hilbert term in its canonical form. We further
put $\kappa =1$. To solve such a gauge condition, an appropriate way
\cite{Kallosh:2000ve} is to change variables from the basis $\{X^I\}$ to
a basis  $\{y,\,z^\alpha\} $, where $\alpha =1,\ldots ,n$ using
\begin{equation}
X^I = y\, Z^I(z)\,.  \label{XyZ}
\end{equation}
We do not specify the $(n+1)$ functions $Z^I$ of the base space
coordinates $z^\alpha $, so that we keep the freedom of arbitrary
coordinates on the base. The $Z^I$ must be non-degenerate in the sense
that the $(n+1)\times (n+1)$ matrix
\begin{equation}
  \begin{pmatrix}
  %\pmatrix{
  Z^I\cr \partial_\alpha Z^I
  %}
  \end{pmatrix}
 \label{nondegenerateZ}
\end{equation}
should have rank $n+1$. There are many ways to choose the $Z^I$. One
simple choice, labeling the $I$ index from $0$ to $n$, can be
\begin{equation}
  Z^0=1\,,\qquad Z^\alpha =z^\alpha \,.
 \label{simplepar}
\end{equation}
Then the gauge condition can be solved for the modulus of $y$. Its phase is
determined by a gauge condition for the $R$-symmetry. The homogeneity
properties then determine that
\begin{equation}
\mathcal{N}= y\,\bar y\, Z^I(z)G_{I\bar J}(z,\bar z)\bar Z^{\bar J}(\bar
z)\,, \qquad G_{I\bar J}(z,\bar z)= \partial_I\partial_{\bar J}\mathcal{N}(X,\bar X)={\cal N}_{I\bar J}\,.  \label{formNhomog}
\end{equation}
The function that acts as K{\"a}hler potential for this gauge is
\begin{equation}
\mathcal{K}(z,\bar z)= -3 \ln\left[ -\ft13 Z^I(z)G_{I\bar
J}(z,\bar z)\bar Z^{\bar J}(\bar z)\right]\,.  \label{calKprojected}
\end{equation}
This defines the K{\"a}hler metric
\begin{equation}
g_{\alpha \bar\beta }=\partial_\alpha \partial_{\bar\beta }\mathcal{K}(z,\bar z)\,.  \label{KahlermetriccalK}
\end{equation}

Note that there is an arbitrariness in the definition (\ref{XyZ}). We may
consider redefinitions
\begin{equation}
y^{\prime}= y \, \rme^{f(z)/3}\,, \qquad Z^{\prime}{}^{\,I}=Z^I\,
\rme^{-f(z)/3}\,.  \label{Kahlertransf}
\end{equation}
These redefinitions lead to a different K{\"a}hler potential:
\begin{equation}
\mathcal{K}^{\prime}(z,\bar z)= \mathcal{K}(z,\bar z)+f(z) + \bar f(\bar
z)\,.  \label{KahlertransfK}
\end{equation}
Hence these can be identified with K{\"a}hler transformations for the K{\"a}hler
potentials defined by (\ref{calKprojected}). In view of these K{\"a}hler
transformations, it is often useful to define K{\"a}hler-covariant
derivatives. The gauge field for the parameter $f(z)$ is then $\partial
_\alpha \mathcal{K}$, while for $\bar f(\bar z)$ it is $\partial
_{\bar\alpha }\mathcal{K}$. In both cases these are the gauge fields
because they transform with a derivative on the parameters. We thus
define
\begin{eqnarray}
\nabla _\alpha Z^I &=& \partial_\alpha Z^I + \ft{1}{3}\left( \partial
_\alpha \mathcal{K}\right) Z^I\,,\qquad \nabla _{\bar \alpha } Z^I =
\partial_{\bar \alpha } Z^I=0\,,  \nonumber\\
\nabla _{\bar \alpha }\bar Z^{\bar I} &=& \partial_{\bar \alpha} \bar
Z^{\bar I} + \ft{1}{3}\left( \partial_{\bar \alpha} \mathcal{K}\right)
Z^I\,,\qquad \nabla _\alpha \bar Z^{\bar I} = \partial_\alpha \bar Z^{\bar
I}=0\,.  \label{Kahlercovder}
\end{eqnarray}
We now define weights of functions under K{\"a}hler transformations. Any
object that transforms like $Z^I$ in (\ref{Kahlertransf}) is defined to
have weights $(w_+,w_-)=(1,0)$.  Hence, $y$ has weight $(-1,0)$. %(at
%least before the gauge fixing of the superconformal $\U(1)$ has been
%taken, see below for the correction after this gauge fixing).

The objects that appear in the superconformal formulation do not
transform under K{\"a}hler transformations. For any quantity that in the
superconformal variables is of the form
\begin{equation}
  {\cal V}(X,\bar X)= y^{w_+}\bar y^{w_-} V(z,\bar z)\,.
 \label{calVV}
\end{equation}
we define that  $V$ has weights $(w_+,w_-)$, and the K{\"a}hler-covariant
derivatives are
\begin{eqnarray}
\nabla _\alpha V & = & \left[\partial_\alpha +\ft13w_+(\partial_\alpha \mathcal{K})\right] V \,, \nonumber\\
  \overline{\nabla} _{\bar \alpha} V  & = & \left[\partial_{\bar \alpha} +\ft13w_-(\partial_{\bar \alpha} \mathcal{K})\right] V\,.
 \label{nablaV}
\end{eqnarray}
Remark that $\Phi $ does not transform under K{\"a}hler transformations, and
thus has weights~(0,0). On the other hand $\rme^{\mathcal{K}/3}$ has
weights $(-1,-1)$, and thus $\nabla _\alpha\rme^{\mathcal{K}/3}=0$.

The gauge and $\U(1)$ transformations on $X^I$ split in those for $y$
and $z^\alpha $ as follows:
\begin{equation}
\delta y= y\left( \ft13\theta^A  r_A(z)+\rmi \lambda _{T}\right) \,,\qquad \delta z^\alpha = \theta^Ak_A^\alpha
(z)\,,  \label{delyzk}
\end{equation}
where $\ft13r_A(z)$ can be considered as the component of the Killing
vectors in the direction of $y$.

\bigskip

{\it Our new setup} will assume the following gauge conditions:
\begin{eqnarray}
D\mbox{-gauge}&\quad &\mathcal{N}=\Phi (z,\bar z)\,, \nonumber\\
U(1)\mbox{-gauge}&\quad &y=\bar y\,,  \label{DU1gauge}
\end{eqnarray}
with an arbitrary function $\Phi(z,\bar z)$. We keep the definition of
$\mathcal{K}$ as in (\ref{calKprojected}), with the associated K{\"a}hler
transformations and covariant derivatives as in (\ref{Kahlercovder}),
and all the above equations remain valid. Furthermore, all the results
below will then reduce to those of \cite{CFGVP-1,Kallosh:2000ve} when
$\Phi =-3$.\footnote{Or $\Phi =-3\kappa^{-2}$, where $\kappa $ is the
gravitational coupling constant which has often been set to~1. To
restore $\kappa $ one also replaces $\exp\mathcal{K}$ with $\kappa
^6\exp \kappa^2\mathcal{K}$, and thus also $g_{\alpha \bar\beta }$ with
$\kappa^2g_{\alpha \bar\beta }$, and $\psi _\mu $ with $\kappa \psi _\mu
$.}

The value for $y$ for the new gauge choice is
\begin{equation}
y=\bar y= \sqrt{-\frac{\Phi }{3}}\exp\frac{\mathcal{K}}{6}\,.  \label{valuey}
\end{equation}
However, in many equations we will keep the phase of $y$ arbitrary. The
$\U(1)$ gauge choice can be taken at any time.

The vanishing of the derivative of the $D$-gauge condition w.r.t. $z^\alpha $
leads to
\begin{equation}
\mathcal{N}_I\nabla_\alpha Z^I =0\,,\qquad \mathcal{N}_I=\bar y\,G_{I\bar
J}\bar Z^{\bar J}\,.  \label{NInablaZ0}
\end{equation}
Note that this equation does not feel the presence of the function
$\Phi$. With these equations one can write the matrix identity
\begin{equation}
\begin{pmatrix}
-3& 0\cr 0 & g_{\alpha \bar\beta }
\end{pmatrix}
=\rme^{\mathcal{K}/3}
\begin{pmatrix}
Z^I\cr \nabla _\alpha Z^I
\end{pmatrix}
G_{I\bar J}
\begin{pmatrix}
\bar Z^{\bar J} & \overline{\nabla} _{\bar\beta }\bar Z^{\bar J}
\end{pmatrix}
\,.  \label{matrixeqngg}
\end{equation}
Every matrix here is $(n+1)\times (n+1)$, and should be invertible.

This matrix identity is useful to translate quantities in the $X^I$ basis
to quantities in the $\{y,z^\alpha\}$ basis. E.g. it implies that the
inverse of $G_{I\bar J}$ is
\begin{equation}
G^{I\bar J}=\rme^{\mathcal{K}/3}\left( -\ft13
Z^I\bar Z^{\bar J} +g^{\alpha \bar\beta }\nabla _\alpha Z^I\,\nabla _{\bar\beta }\bar Z^{\bar J}\right) \,.  \label{inverseG}
\end{equation}

We assume that $\Phi $ is a (Yang-Mills) gauge-invariant function. Hence
the dilatation gauge condition is invariant. However, the $\U(1)$-gauge
is not, see the transformations (\ref{delyzk}), and it is not invariant
under K{\"a}hler transformations (\ref{Kahlertransf}) either. It is not
invariant under supersymmetry either, but we postpone this for when we
have discussed a new basis of the fermions. This implies that we cannot
forget the transformations with parameter $ \lambda _{T}$, but should
relate it to the gauge transformations and K{\"a}hler transformations (and
later also to supersymmetry)
\begin{equation}
  \lambda _{T}= \ft16\rmi \theta^A\left[  r_A -\bar r_A \right] +\ft16\rmi\left[ f(z) -\bar f(\bar z)\right]\,.
 \label{decompU1bos}
\end{equation}

Taking this into account, and also the gauge invariance of $\Phi $, we
find that the K{\"a}hler potential $\mathcal{K}$ transforms under gauge
transformations as
\begin{equation}
\delta \mathcal{K}=\,\theta^A \left[r_A(z)+\bar r_A(\bar z)\right] \,.
\label{delcalKA}
\end{equation}
The moment map ${\cal P}_A$ defined in (\ref{eqnskconform}) depends on
this quantity $r_A(z)$ as
\begin{equation}
\mathcal{P}_A=(-\ft13\Phi )P_A\,,\qquad P_A=\rmi\left( k_A{}^\alpha \partial_\alpha \mathcal{K}%
-\,r_A\right)= -\rmi\left( k_A{}^{\bar\alpha} \partial_{\bar\alpha}
\mathcal{K}-\,\bar r_A\right) \,.  \label{momentmapfromr}
\end{equation}
Another convenient way to state this, is to write the Killing vectors in
in the $X^I$ basis as
\begin{equation}
k_A^I = y\left[ k_A^\alpha \nabla _\alpha Z^I +\ft13\rmi P_AZ^I\right] \,.  \label{Killingrelembproj}
\end{equation}

The bosonic part of the value of the auxiliary field $\mathcal{A}_\mu $,
see (\ref{fullAmu}) is
\begin{eqnarray}
   \mathcal{A}_\mu &=& \ft16\rmi\left(
\partial_\mu z^\alpha \partial_\alpha \mathcal{K}
- \partial_\mu \bar z^{\bar \alpha }\partial_{\bar \alpha }\mathcal{K}\right)
-\ft13 A_\mu {}^AP_A \nonumber\\
&=&\ft16\rmi\,
  \left( \hat\partial_\mu z^\alpha\partial_\alpha\mathcal{K}
  - \hat\partial_\mu \bar z^{\bar \alpha }\partial_{\bar \alpha }\mathcal{K}\right)+\ft16\rmi A_\mu^A(r_A-\bar r_A) \,.
 \label{valueAmuN1}
\end{eqnarray}
Independent of the gauge conditions, one proves that the kinetic terms
of the scalars, $\mathcal{L}_0$ in (\ref{confactkin}) is
\begin{eqnarray}
  \mathcal{L}_0&=&-\frac1{4\mathcal{N}}(\partial_\mu \mathcal{N})(\partial^\mu  \mathcal{N})
  -\mathcal{N}(\hat\partial_\mu z^\alpha )\,(\hat\partial^\mu\bar z^{\bar\beta })\frac{\partial}{\partial z^\alpha}\frac{\partial}
  {\partial\bar z^{\bar\beta }}
  \ln \left[ Z^I(z)G_{I{\bar J}}\bar Z^{{\bar J}}(\bar z)\right]
  \,, \nonumber\\
\hat\partial_\mu z^\alpha &\equiv &\partial_\mu z^\alpha -A_\mu^A
k_A{}^\alpha\,.
 \label{LagrdKdz}
\end{eqnarray}
After the gauge choice, this is thus
\begin{eqnarray}
  \mathcal{L}_0&=& -\frac1{4\Phi }(\partial_\mu \Phi )(\partial^\mu \Phi )
  +\ft13g_{\alpha \bar\beta }\Phi (\hat\partial_\mu z^\alpha )\,(\hat\partial^\mu\bar z^{\bar\beta })
  \,,\nonumber\\
  &=&\Phi\left(\ft13 g_{\alpha \bar\beta }-\ft12L_\alpha L_{\bar\beta }\right)(\hat\partial_\mu z^\alpha )\,(\hat\partial^\mu\bar z^{\bar\beta })
  -\ft14\Phi\left[  L_\alpha L_\beta(\hat\partial_\mu z^\alpha )\,(\hat\partial^\mu z^{\beta })+\hc\right] \,,
 \label{L0Phi}
\end{eqnarray}
which is the same as (\ref{Turin-5}), and where we introduced $L$ for
$\ln\Phi $ and
\begin{equation}
  L_\alpha = \partial_\alpha\ln (-\Phi)\,,\qquad L_{\bar\alpha} = \partial_{\bar\alpha}\ln (-\Phi)\,.
 \label{Lalpha}
\end{equation}

For the superpotential, we define $\mathcal{W}(X)=y^3 W(z)$. Hence
$W(z)$ has K{\"a}hler weights (3,0). This leads to
\begin{equation}
\mathcal{W}_IZ^I = 3y^2W(z)\,,\qquad y^{-2}\mathcal{W}_I\nabla _\alpha Z^I =
\nabla _\alpha W\equiv \partial_\alpha W +(\partial_\alpha \mathcal{K})W\,.
\label{identWcalW}
\end{equation}
The $F$-term in the superpotential therefore reduces to (taking only bosonic
terms from the field equation of $F^I$)
\begin{equation}
V_F=%&=& \rme^{\mathcal{K}/3}y^4\left(-3W \overline{W}
%+ \nabla _\alpha W g^{\alpha \bar\beta }\nabla _{\bar\beta }\overline{W}\right)\nonumber\\ &=&
\ft19\Phi^2\rme^{\mathcal{K}}\left(-3W \overline{W} + \nabla _\alpha W g^{\alpha \bar\beta }\nabla _{\bar\beta }\overline{W}\right)\,.  \label{VF}
\end{equation}
This agrees with what we already expected in (\ref{Turin-3}). Due to
(\ref{momentmapfromr}), also the $D$-term has the same overall
$\Phi$-dependence
\begin{equation}
V_D=\ft1{18}\Phi^2(\Re f)^{-1\,AB} P_AP_B\,,  \label{VD}
\end{equation}
This agrees with what we found in (\ref{Turin-3}).

We introduce now modified K{\"a}hler-covariant derivatives, which take the
presence of $\Phi$ into account. For an object $V$ that has weights
$(w_+,w_-)$, we define
\begin{eqnarray}
 \widetilde \nabla _\alpha V & = & \left[\partial_\alpha +\ft13w_+(\partial_\alpha \mathcal{K})+\ft12(w_++w_-)L_\alpha\right] V \,, \nonumber\\
  \widetilde {\overline{\nabla}} _{\bar \alpha} V  & = & \left[\partial_{\bar \alpha}
  +\ft13w_-(\partial_{\bar \alpha} \mathcal{K})+\ft12(w_++w_-)L_{\bar \alpha}
  \right] V\,.
 \label{tildenablaV}
\end{eqnarray}
We can also define the covariant derivatives in spacetime, using
\begin{equation}
  \widetilde \nabla _\mu=(\hat{\partial} _\mu z^\alpha)\widetilde{\nabla}  _\alpha +(\hat{\partial} _\mu \bar z^{\bar \alpha})
  \widetilde{\overline{\nabla}}  _{\bar \alpha}\,.
 \label{nablamu}
\end{equation}
One can evaluate these before or after gauge fixing of the $\U(1)$
symmetry. Before the latter is gauge-fixed we have to add the
covariantization of the latter. $y$ has then weight $(-1,0)$, so that we
define
\begin{equation}
  \widetilde{\nabla} _\mu y= \left( \partial_\mu -\rmi \mathcal{A}_\mu -\ft13\hat{\partial} _\mu z^\alpha \partial_\alpha \mathcal{K}
  +\ft12\partial_\mu L -\ft13 A_\mu^A r_A\right) y = 0\,.
 \label{nablay}
\end{equation}
The calculation is modified after the $\U(1)$ gauge fixing, but the
result is still the same. The $\U(1)$ transformation is gone, but due to
(\ref{decompU1bos}) and (\ref{delyzk}), the  K{\"a}hler transformation of
$y$ is (in agreement with its value in (\ref{valuey}))
\begin{equation}
  y'= y \exp \left[f(z)+\bar f(\bar z)\right]/6\,.
 \label{Kahleryafter}
\end{equation}
Thus $y$ has now K{\"a}hler weights $(w_+,w_-)=(-\ft12,-\ft12)$, leading
again to
\begin{equation}
  \widetilde{\nabla} _\mu y= \left( \partial_\mu  -\ft16\hat{\partial} _\mu z^\alpha \partial_\alpha \mathcal{K}
  -\ft16\hat{\partial} _\mu \bar z^{\bar \alpha} \partial_{\bar\alpha} \mathcal{K}
  +\ft12\partial_\mu L -\ft16A_\mu^A(r_A+\bar r_A)\right) y = 0\,.
 \label{nablaygf}
\end{equation}
We thus find that $y$ is invariant under the new covariant derivatives
\begin{equation}
 \widetilde \nabla _\alpha y=\widetilde {\overline{\nabla}} _{\bar \alpha} y=0\,.
 \label{tildenaby}
\end{equation}
This will facilitate many calculations.

There are some differences between these modified covariant derivatives
and the ordinary covariant derivatives (\ref{Kahlercovder}). Most
important is that the anti-chiral modified covariant derivative does not
vanish on $Z^I$:
\begin{equation}
  \widetilde{\overline\nabla} _{\bar\alpha} Z^I=
\ft12Z^I  L_{\bar \alpha }\,.
 \label{bnabZ}
\end{equation}

The commutator of the covariant derivatives on scalar functions still
satisfies the rule
\begin{equation}
  \left[ \widetilde{\nabla} _\alpha ,\widetilde{\overline\nabla} _{\bar\beta }\right] V(z,\bar z)=
  \ft13(w_--w_+)g_{\alpha \bar\beta }V(z,\bar z)\,.
 \label{commnablaKahlerw}
\end{equation}
This leads also to an expression that we will need below:
\begin{equation}
  \widetilde{\overline{\nabla }}_{\bar\beta }\widetilde{\nabla }_\alpha Z^I= Z^I\left(\ft13g_{\alpha \bar\beta }
  +\ft12 L_{\alpha\bar\beta }
  \right)+\ft12L _{\bar\beta }\widetilde{\nabla }_\alpha Z^I \,.
 \label{nbnZ}
\end{equation}

The matrix equation (\ref{matrixeqngg}) gets modified:
\begin{equation}
\begin{pmatrix}
\Phi & \ft12 \Phi L_{\bar\beta } \cr \ft12  \Phi L_\alpha
&\tilde g_{\alpha \bar\beta }
\end{pmatrix}
=y\bar y
\begin{pmatrix}
Z^I\cr \widetilde{\nabla} _\alpha Z^I
\end{pmatrix}
G_{I\bar J}
\begin{pmatrix}
\bar Z^{\bar J} & \widetilde{\overline{\nabla}} _{\bar\beta }\bar Z^{\bar J}
\end{pmatrix}
\,.  \label{matrixeqnggtildeb}
\end{equation}
where
\begin{equation}
  \tilde g_{\alpha\bar\beta }\equiv -\ft13\Phi g_{\alpha\bar\beta }+\ft14\Phi L _\alpha L _{\bar\beta }\,.
 \label{deftildeg}
\end{equation}

%The inverse of the matrix at the left-hand side is
%\begin{equation}\frac{-\Phi }{3}
%\begin{pmatrix}-\frac13+\frac{|\partial\Phi |^2}{4\Phi^2}&-\frac{1}{2}g^{\alpha \bar \alpha }\partial_{\bar \alpha }L \cr
%-\frac{1}{2}g^{\bar\beta \beta  }\partial_{\beta }L &g^{\alpha \bar\beta }\end{pmatrix}\,.
%\end{equation}

To obtain the second holomorphic derivative of $Z^I$, one can take a
covariant derivative on the second line of (\ref{matrixeqnggtildeb}) to
obtain
\begin{equation}
  \widetilde{\nabla} _\beta \widetilde{\nabla} _\alpha Z^I=-y\, \Gamma^I_{JK}\widetilde{\nabla} _\alpha Z^J\widetilde{\nabla} _\beta Z^K
  +L_{(\alpha }\widetilde{\nabla }_{\beta)} Z^I+Z^I\left( \ft12L _{\alpha \beta }  -\ft14L _\alpha L _\beta \right)  \,,
 \label{tnabtnabZ}
\end{equation}
where
\begin{equation}
  L _{\alpha\beta }=\nabla _\alpha L_\beta =
  \partial_\alpha L_\beta -\Gamma _{\alpha\beta }^\gamma L _\gamma \,.
 \label{Lalphabeta}
\end{equation}
This can be used further to calculate the curvature of the projective
manifold. Indeed, acting with $y\bar y\widetilde{\overline\nabla}
_{\bar\beta }\bar Z^{\bar J}G_{I\bar J}\widetilde{\overline\nabla}
_{\bar \alpha}$ on this equation, and using that on a vector quantity
\begin{equation}
  \left[\widetilde{\overline\nabla} _{\bar\alpha  },\, \widetilde{\nabla} _\beta \right]\widetilde{\nabla}
   _\alpha Z^I=
  \ft13g_{\beta \bar\alpha  }\widetilde{\nabla }_\alpha Z^I+R_{\bar \alpha\beta \alpha }{}^\gamma \widetilde{\nabla }_\gamma
  Z^I\,,
 \label{commnablavector}
\end{equation}
we obtain after many cancellations of $L$-dependent terms
\begin{equation}
(-\ft13\Phi ) \left[  R_{\alpha\bar \alpha\beta \bar\beta }-\ft23 g_{\bar \alpha (\alpha }g_{\beta )\bar\beta }\right] =
(y\bar y)^2 R_{I\bar IJ\bar J} \widetilde{\nabla }_\alpha Z^I\widetilde{\nabla} _\beta Z^J\widetilde{\overline\nabla} _{\bar\alpha  }\bar Z^{\bar I}
\widetilde{\overline\nabla} _{\bar\beta }\bar Z^{\bar J}\,.
 \label{curvrelation}
\end{equation}
Observe that the cancellations can be explained due to the fact that the
dilatational symmetry of the embedding manifold implies that $Z^I
R_{I\bar IJ\bar J}=0$.

\subsection{The physical fermions}

In order to define the physical bosons in the previous section, we
changed from the conformal basis $\{X^I\}$ to the basis $\{y,z^\alpha
\}$. We now make a similar change of basis from the conformal fermions
$\{\Omega^I\}$ to a new basis\footnote{We again use the implicit chiral
notation, i.e. $P_L\chi^\alpha =\chi^\alpha $ and $P_R\chi^{\bar\alpha
}=\chi^{\bar \alpha }$.} $\{\chi^0,\chi^\alpha \}$, using
\begin{eqnarray}
\Omega^I &=&y \,\left( \chi^0Z^I+ \chi^\alpha \widetilde{ \nabla}_\alpha
Z^I \right) \,.  \label{defchi0alpha}
\end{eqnarray}
Our aim is to have $\chi^0=0$ as gauge condition for the $S$-gauge
transformations. We therefore choose
\begin{equation}
S\mbox{-gauge}\quad \mathcal{N}_I \Omega^I = \ft12\Phi_\alpha \chi^\alpha \,,
\label{Sgauge}
\end{equation}
which is equivalent to $\chi^0=0$. Hence the gauge fixed fermions are
$\Omega^I= y\chi^\alpha \widetilde{ \nabla}_\alpha Z^I$.

The covariant derivative of the physical fermion is
\begin{equation}
  \PLa  D_\mu \chi^\alpha  =  \PLa\left(\partial_\mu+\ft14\omega _\mu {}^{ab}(e)\gamma _{ab}+\ft32\rmi\mathcal{A}_\mu  \right) \chi^\alpha -A_\mu^A
  \frac{\partial k_A{}^\alpha (z)}{\partial
  z^\beta }\PLa\chi
^\beta + \Gamma^\alpha _{\beta \gamma }
\PLa\chi^\gamma \hat\partial_\mu z^\beta \,.
 \label{hatD0chi}
\end{equation}
The covariant derivative on the conformal fermions (\ref{covderconf})
can then be rewritten as
\begin{eqnarray}
   \hat{D}_\mu \PLa\Omega^I&=& D_\mu \PLa\left(y\chi^\alpha\widetilde{\nabla}_\alpha Z^I\right)
    +y^2\Gamma^I_{JK}\chi^\alpha \widetilde{\nabla}_\alpha Z^K \widetilde{\nabla}_\beta Z^J \hat{\partial}_\mu z^\beta \nonumber\\
    &=& y\left(D_\mu \PLa\chi^\alpha\right) \widetilde{\nabla}_\alpha Z^I +\PLa y\chi^\alpha
     \left(\hat{\partial} _\mu z^\beta  \widetilde{\nabla} _\beta  +\hat{\partial} _\mu \bar z^{\bar\beta }
      \widetilde{\overline{\nabla}} _{\bar\beta }\right)\widetilde{\nabla}_\alpha Z^I+y^2\Gamma^I_{JK}\chi^\alpha \widetilde{\nabla}_\alpha Z^K \widetilde{\nabla}_\beta Z^J \hat{\partial}_\mu z^\beta\nonumber\\
 &=& y\left(D_\mu \PLa\chi^\alpha\right) \widetilde{\nabla}_\alpha Z^I+\PLa y\chi^\alpha
    \hat{\partial} _\mu z^\beta \left[L_{(\alpha }\widetilde{\nabla }_{\beta)} Z^I
    +Z^I\left( \ft12L _{\alpha \beta }  -\ft14L _\alpha L _\beta\right) \right]
    \nonumber\\
    && +\PLa y\chi^\alpha\hat{\partial} _\mu \bar z^{\bar\beta }\left[Z^I\left(\ft13g_{\alpha \bar\beta }
  +\ft12 L_{\alpha \bar\beta }
  \right)+\ft12L _{\bar\beta }\widetilde{\nabla }_\alpha Z^I  \right]  \,,
 \label{translateDchi}
\end{eqnarray}
using (\ref{tnabtnabZ}) and (\ref{nbnZ}). This can be inserted in the
kinetic fermion terms, $\mathcal{L}_{1/2}$ in (\ref{confactkin}). The
contribution of the last line of (\ref{translateDchi}) can be complex
conjugated such that this leads to
\begin{equation}
 \mathcal{L}_{1/2}=-\ft12\tilde g_{\alpha\bar\beta }
   \bar \chi^{\bar\beta } \PRa \slashed{D} \chi^\alpha
+\ft12\Phi  \bar\chi^\alpha\PLa\gamma^\mu \chi^{\bar\beta }\hat{\partial}_\mu z^\gamma \left[-\ft13 g_{\gamma \bar\beta }L_\alpha+\ft14L _{\alpha\gamma }
L _{\bar\beta }-\ft14L_\alpha L _{\gamma\bar\beta}\right]+\hc\,.
 \label{L12step4}
\end{equation}

Now we consider the fermion mass terms, $\mathcal{L}_m$ in
(\ref{confact2fermion}). We rewrite them as
\begin{equation}
\mathcal{L}_m  =   \ft12 m_{3/2}\bar \psi _\mu P_R
\gamma^{\mu\nu}\psi _\nu  -\ft12 m_{\alpha\beta }\bar
\chi^\alpha\PLa\chi^\beta
-m_{\alpha A}\bar \chi^\alpha\PLa\lambda^A-\ft12m_{AB}\bar\lambda^AP_L\lambda^B +\hc\,, \label{fermionmass}
\end{equation}
where the (complex) gravitino mass parameter can be easily recognized as
\begin{equation}
   m_{3/2}={\cal  W}= y^3W= \left( -\ft13\Phi \right)^{3/2} \rme^{\mathcal{K}/2}W\,.
 \label{m32N1}
\end{equation}
For the mass terms of the chiral fermions, we have
\begin{equation}
  m_{\alpha \beta }\chi^\alpha \PLa\chi^\beta=\nabla _I{\cal  W}_J\bar \Omega^I\PLa\Omega^J\,.
 \label{malbe1}
\end{equation}
We first observe that in (\ref{identWcalW}) we can insert as well the
modified covariant derivatives $\widetilde{\nabla }$ due to the
homogeneity conditions. Then we take a further covariant derivative
using again (\ref{tnabtnabZ}), gives
\begin{equation}
 \tilde \nabla _\beta \tilde \nabla _\alpha W= y^{-1}\nabla _J{\cal  W}_I\widetilde{\nabla }_\beta Z^J
 \widetilde{\nabla }_\alpha  Z^I +y^{-2}{\cal W}_I\left(L _{(\alpha} \widetilde{\nabla }_{\beta)} Z^I
+\ft12L _{\alpha \beta }Z^I
  -\ft14L _\alpha L _\beta Z^I\right)\,.
 \label{tnabtnabW}
\end{equation}
Therefore,
\begin{eqnarray}
m_{\alpha \beta }&=&  y^2\nabla _I{\cal  W}_J\widetilde{\nabla }_\beta Z^J\widetilde{\nabla }_\alpha  Z^I
\nonumber\\ &=&
y^3\widetilde{\nabla }_\beta \widetilde{\nabla }_\alpha W
-y{\cal W}_I\left(L _{(\alpha} \widetilde{\nabla }_{\beta)} Z^I
+\ft12L _{\alpha \beta }Z^I
  -\ft14L _\alpha L _\beta Z^I\right)
\nonumber\\
&=&
y^3\left[ \widetilde{\nabla }_\beta \widetilde{\nabla }_\alpha W
-L _{(\alpha} \widetilde{\nabla }_{\beta)} W
-3W\left(\ft12L _{\alpha \beta }-\ft14L _\alpha L _\beta \right)\right] \nonumber\\
&=&\left( -\ft13\Phi \right)^{3/2} \rme^{\mathcal{K}/2}\left[\nabla _\beta \nabla _\alpha W
+2L _{(\alpha} \nabla _{\beta)} W
 \right] \,.
 \label{nabWI}
\end{eqnarray}
For the mass terms involving $\lambda $, we first need an equation for
the derivative of $f_{AB}$:
\begin{equation}
  f_{AB\,\alpha }=\nabla _\alpha f_{AB}= y\ f_{AB\,I}\widetilde{\nabla } _\alpha Z^I= y\ f_{AB\,I}\nabla_\alpha Z^I\,.
 \label{fABalpha}
\end{equation}
A further derivative on this equation is relevant for the 4-fermion
terms. Using (\ref{tnabtnabZ}) and the homogeneity of degree zero of
$f_{AB}$ so that $f_{AB\, I}Z^I=0$, we obtain
\begin{equation}
\nabla _\beta f_{AB\,\alpha }=\widetilde{\nabla }_\beta f_{AB\,\alpha }=y^2\nabla _Jf_{AB\,I}\widetilde{\nabla }_\alpha Z^I
\widetilde{\nabla }_\beta Z^J
+L _{(\alpha} \partial_{\beta )}f_{AB} \,.
 \label{finalident}
\end{equation}

For the $\lambda \lambda $ mass term we use the expression for $G^{I\bar
J}$ in (\ref{inverseG}), the same homogeneity equation of $f_{AB}$,
(\ref{identWcalW}) and (\ref{fABalpha}) to translate
\begin{equation}
 m_{AB}=-\ft12 G^{I\bar J}\overline{{\cal W}}_{\bar J} f_{AB I}
  =-\ft12(-\ft13\Phi)^{1/2}\rme^{\mathcal{K}/2}f_{AB\,\alpha } g^{\alpha \bar\beta }\overline\nabla_{\bar\beta } \overline{W}\,.
 \label{fermmassN1SUGRA}
\end{equation}
The conformal expression of the $\lambda \chi $ mass term, gives
\begin{equation}
  m_{\alpha A}=
  \rmi\sqrt{2}\left[ \partial_I{\cal P}_A-\ft1{4}f_{ABI}(\Re f)^{-1\,BC} {\cal P}_C\right]y\widetilde{\nabla }_\alpha Z^I\,.
 \label{confmixedm}
\end{equation}
We first calculate
\begin{eqnarray}
  \partial_\alpha {\cal P}_A&=&y\widetilde{\nabla } _\alpha Z^I\partial_I{\cal P}_A + \bar y\widetilde{\nabla } _\alpha \bar Z^{\bar I}
  \partial_{\bar I}{\cal P}_A\nonumber\\
  &=&y\widetilde{\nabla } _\alpha Z^I\partial_I{\cal P}_A +\ft12L_\alpha {\cal P}_A\,,
 \label{calcdPA}
\end{eqnarray}
due to (\ref{bnabZ}) and the homogeneity equation $X^{\bar I}\partial
_{\bar I}{\cal P}_A={\cal P}_A$. Using (\ref{momentmapfromr}) this gives
\begin{equation}
  y\widetilde{\nabla } _\alpha Z^I\partial_I{\cal P}_A =-\ft13\Phi \partial_\alpha P_A-\ft16P_A\partial_\alpha \Phi \,.
 \label{firstpartmalphaA}
\end{equation}
Using also again (\ref{fABalpha}), we obtain
\begin{equation}
   m_{\alpha A}=
  -\ft13\rmi\sqrt{2}\Phi\left[  \left( \partial_\alpha +\ft12L_\alpha \right)P_A-\ft1{4} f_{AB\,\alpha }(\Re f)^{-1\,BC} P_C\right]\,.
 \label{malphaA}
\end{equation}

For $\mathcal{L}_d$ in (\ref{confact2fermion}) we need only one new
calculation:
\begin{eqnarray}
   G_{I\bar J}D_\mu \bar X^{\bar J}y\widetilde{\nabla }_\alpha Z^I&  = &y\bar yG_{I\bar J}
 \left( \hat{\partial}_\mu z^\beta \widetilde \nabla _\beta +\hat{\partial}_\mu \bar z^{\bar\beta } \widetilde {\overline{\nabla}} _{\bar\beta} \right)\bar Z^{\bar J} \widetilde{\nabla }_\alpha Z^I   \nonumber\\
   & = & \ft14\Phi L _\alpha L _\beta \hat{\partial}_\mu z^\beta
   +\tilde g_{\alpha \bar\beta }\hat{\partial}_\mu \bar z^{\bar\beta }\,.
 \label{calcLd}
\end{eqnarray}
To calculate the 4-fermion terms, we need the fermionic part of the
auxiliary field $A_\mu $. Its conformal expression was given in
(\ref{fullAmu}), which can be evaluated as
\begin{equation}
  A_\mu^{\mathrm{F}}=\frac{\rmi}{4\sqrt{2}}\bar \psi_\mu\left(L_\alpha
\PLa\chi^\alpha -L_{\bar\alpha}
 \PRa \chi^{\bar \alpha }\right)  +\frac{\rmi}{4\Phi}\tilde g_{\alpha \bar\beta }\bar \chi^\alpha  \PLa\gamma _\mu \chi
^{\bar\beta }
+\frac{3\rmi}{8\Phi } (\Re f_{A B }) \bar\lambda^A \gamma _\mu \gamfive\lambda^B\,.
 \label{AmuF}
\end{equation}
 % The square of these contain various terms. The $(\psi _\mu \chi)^2$
%terms combine with a term in $\mathcal{L}_{\rm 4f}$ that is of the form
%$-\ft12\tilde g_{\alpha \bar\beta }\bar \psi _\mu \PRa\chi^{\bar\beta
%}\,\bar \psi _\mu \PLa\chi^\alpha$.

One term in the square of this expression is the $\chi^4$ term, which
combines (after a Fierz transformation) with the curvature term in
$\mathcal{L}_{\mathrm{4f}}$, where (\ref{curvrelation}) is now
convenient. The result is given in the beginning of the paper, in Sec.
\ref{ss:N1Jordan}. Here we still give the 4-fermion term:
\begin{eqnarray}
 \mathcal{L}_{4\mathrm{f}}&=&\frac{1}{96}\Phi \left[ (\bar{\psi}
^\rho\gamma^\mu\psi^\nu) ( \bar{\psi}_\rho\gamma_\mu\psi_\nu +2 \bar{\psi}
_\rho\gamma_\nu\psi_\mu) - 4 (\bar{\psi}_\mu \gamma\cdot\psi)(\bar{\psi}^\mu
\gamma\cdot\psi)\right]\nonumber\\
&&+\left\{  - \frac1{4\sqrt{2}}f_{A B\,\alpha }\bar \psi \cdot\gamma  \chi^\alpha  \bar\lambda^A P_L\lambda^B
 +\frac18 \bar \chi^\alpha \PLa\chi^\beta  \bar\lambda^AP_L
 \lambda^B\left[
\nabla _\beta f_{AB\,\alpha }-L
_\alpha f_{AB\,\beta }
 \right]  + \hc \right\}\nonumber\\
 &&+ \ft{1}{16}\edet^{-1}\varepsilon^{\mu\nu\rho\sigma} \bar
\psi _\mu \gamma_\nu \psi _\rho
\left(\ft12\rmi\Re f_{AB}\bar\lambda^A\gamfive\gamma _\sigma \lambda^B
+\tilde g_{\alpha \bar\beta }\bar \chi^{\bar\beta }\PRa\gamma _\sigma \chi^\alpha \right)\nonumber\\
&&+\ft16\Phi  g_{\alpha \bar\beta }\bar \psi _\mu \PRa\chi^{\bar\beta }\,\bar \psi^\mu \PLa\chi^\alpha
 -\ft1{32}\Phi\bar \psi _\mu\left( \chi^\alpha L_\alpha +\chi^{\bar \alpha }L_{\bar \alpha }\right)
\bar \psi^\mu\left( \chi^\beta L_\beta +\chi^{\bar\beta }L_{\bar\beta }\right) \nonumber\\
&& -\frac 9{64\Phi }\left[ (\Re f_{AB })
\bar\lambda^A  \gamma _\mu \gamfive\lambda^B\right]^2 +
+\frac{3}{16}g^{\alpha \bar\beta }\Phi^{-1} f_{A B \,\alpha }\bar\lambda^A P_L\lambda^B
 \bar f_{CD\,\bar\beta }\bar\lambda^C P_R\lambda^D\nonumber\\
&&
 +\ft1{16}(\Re f)^{-1\,AB}
\left( f_{AC\,\alpha }\bar \chi^\alpha \PLa - \bar f_{AC\,\bar\alpha}
\bar \chi^{\bar \alpha }\PRa \right)\lambda^C \left( f_{BD\,\beta }\bar \chi^\beta   \PLa-\bar f_{BD\,\bar\beta }
\bar \chi^{\bar\beta }\PRa\right) \lambda^D \nonumber\\
  &&
\left(-\ft14 g_{\alpha \bar\beta }+\ft3{16}L _\alpha L _{\bar\beta } \right) (\Re f_{AB})\bar \chi^\alpha \lambda^A\bar \chi^{\bar\beta }
\lambda^B\nonumber\\
&&-\frac{1}{8\sqrt{2}}\left( g_{\alpha \bar\beta }\bar \chi^\alpha \gamma^\mu \chi^{\bar\beta }
+\frac{3}{2}(\Re f_{A B }) \bar\lambda^A \gamma _\mu \gamfive\lambda^B\right)
\bar \psi _\mu \left( L_\gamma \chi^\gamma -L_{\bar \gamma }\chi^{\bar
\gamma }\right)\nonumber\\
&&  -\ft1{12}\Phi \left(R_{\alpha \bar  \gamma \beta\bar \delta }-\ft12\kappa^2
g_{\alpha \bar \gamma }g_{\beta \bar \delta }-\ft14L_\alpha L_{\bar \gamma}g_{\beta \bar \delta }+\ft3{32}L_\alpha
L_\beta  L_{\bar \gamma}L_{\bar \delta }\right)\bar \chi^\alpha \PLa\chi
^\beta \bar \chi^{\bar \gamma }\PRa\chi^{\bar \delta }\,.
 \label{4fermionfinal}
\end{eqnarray}

 %This leads to a term
%\begin{equation}
%  -\ft1{12}\Phi \left(R_{\alpha \bar  \gamma \beta\bar \delta }-\ft12\kappa^2
%g_{\alpha \bar \gamma }g_{\beta \bar \delta }-\ft14L_\alpha L_{\bar \gamma}g_{\beta \bar \delta }+\ft3{32}L_\alpha
%L_\beta  L_{\bar \gamma}L_{\bar \delta }\right)\bar \chi^\alpha \PLa\chi
%^\beta \bar \chi^{\bar \gamma }\PRa\chi^{\bar \delta }\,.
% \label{4chiterm}
%\end{equation}

\section{Conclusions}

The main goal of our paper was to derive a complete formulation of
$N=1$, $ d=4$ supergravity in a generic Jordan frame. We found that, in
general, this formulation is very non-trivial. It involves modified
{K{\"a}hler geometry (in the sense specified in our treatment), and it gives
rise to many new complicated terms in the supergravity Lagrangian.}

However, we identified a subclass of theories where the resulting
formulation is remarkably simple. This subclass includes the recently
proposed model of Einhorn and Jones \cite{Einhorn:2009bh}, which was
introduced as an $N=1$ supergravity realization of the Higgs field
inflation \cite{Sha-1}. We found that the inflationary regime in this
model is unstable.

Hopefully, however, the general formalism developed in our paper may
allow one to find new realistic inflationary models in supergravity. As
a starting approach, one can simply study in the Jordan frame several
classes of inflationary models in supergravity, which were found long
time ago in the Einstein frame. As shown by the example of the Higgs
inflation, sometimes it is helpful to identify and study various
physical features of the cosmological models by switching from one frame
to another.

\section*{Acknowledgments}

We are grateful to S. Dimopoulos, M. Einhorn, D. Freedman, P. Graham, R.
Harnik, T. Jones, L. Kofman, S. Mukohyama, S. Shenker, L. Susskind, A.
Westphal for the useful discussions. The work of RK and AL is supported
by the NSF grant 0756174. The work of SF is supported by ERC Advanced
Grant n.226455, \textit{Supersymmetry, Quantum Gravity and Gauge Fields}
(\textit{Superfields}), in part by PRIN 2007-0240045 of Torino
Politecnico, in part by DOE Grant DE-FG03-91ER40662 and in part by INFN,
sez. L.N.F. The work of AM is supported by an INFN visiting Theoretical
Fellowship at SITP, Stanford University, Stanford, CA, USA. The work of
AVP is supported in part by the FWO - Vlaanderen, project G.0235.05, and
in part by the Federal Office for Scientific, Technical and Cultural
Affairs through the `Interuniversity Attraction Poles Programme --
Belgian Science Policy' P6/11-P.

\appendix

\section{Stability with respect to the angle $\protect\beta$}

As we found in Subsecs. \ref{Steste} and \ref{Steste2}, the inflationary
trajectory with $s=0$ (\ref{infl}) of the Einhorn-Jones model
\cite{Einhorn:2009bh} is unstable with respect to a rapid generation of
the $s$ field. Other scalar fields may also have nontrivial dynamical
properties. If after a modification of this model one can find a way to
stabilize the $s$ field, one would then need to study the cosmological
behavior of all other fields. As an example, in this Appendix we will
analyze the behavior of the angle $\beta $, ignoring the issue of the
$s$ field instability.

For $h^2\ll 1$ (consistent with (\ref{infl})), the Einstein frame
potential $V_{E}$ of the fields $h$ and $\beta $ reads
\begin{equation}
V_{E}(h,\beta )={\frac{2h^{4}\lambda^2\sin^22\beta
+(g^2+g^{\prime }{}^2)h^{4}\cos^22\beta }{2(2+h^2\chi \sin
2\beta )^2}}\,.
\end{equation}
The first term in the numerator originates from the $F$-term, the second
term from the $D$-term.

During inflation, in the slow roll regime at $h^2\chi \gg 1$ (see
(\ref{infl})), the potential with respect to $\beta $ is minimized by
the condition of $D$-flatness, corresponding to $\beta =\pi /4$. In this
regime,
\begin{equation}
V_{E}(h,\beta =\pi /4)={\frac{h^{4}\lambda^2}{(2+h^2\chi
)^2}}\approx {\frac{\lambda^2}{\chi^2}}\,.
\end{equation}
One could contemplate the possibility of an additional slow-roll regime
with respect to the slow variation of $\beta$ \cite{Einhorn:2009bh}.
However, the stabilization of $\beta$ during inflation is very firm.
Indeed, when we take into account that the non-canonical kinetic terms
in the angular direction near the minimum are proportional to
${\frac{1}{\chi }}$, we find that the effective mass squared of the
fluctuations of the field $\beta$ is given by
\begin{equation}
m_\beta^2\sim {\frac{h^2\left[ -4\lambda
^2+(g^2+g^{\prime }{}^2)(2+h^2\chi )\right] }{(2+h^2\chi
)^2}}\,.\label{Sste}
\end{equation}
During inflation, in the limit $\chi h^2\gg 1$
\begin{equation}
m_\beta^2={\frac{g^2+g^{\prime }{}^2}{\chi }}\,,
\end{equation}
and the slow-roll parameter $\eta $ with respect to the field $\beta $
thus reads
\begin{equation}
\eta _{_\beta }\equiv {\frac{m_\beta^2}{V}}\approx \chi \,{\frac{
g^2+g^{\prime }{}^2}{\lambda^2}}\,.\label{Ssste}
\end{equation}
This means that for $\chi (g^2+g^{\prime }{}^2)\gg \lambda^2$, one has
$\eta _{_\beta }\gg 1$. Thus, there is no slow-roll regime with respect
to the change of $\beta $ during inflation, because the mass squared of
perturbations of the angle $\beta $ is much greater than $H^2$.
Therefore, {\it during inflation} the field $\beta $ rapidly approaches
$\pi /4$ and stays there.

\begin{figure}[h]
\centering
\includegraphics[scale=0.35]{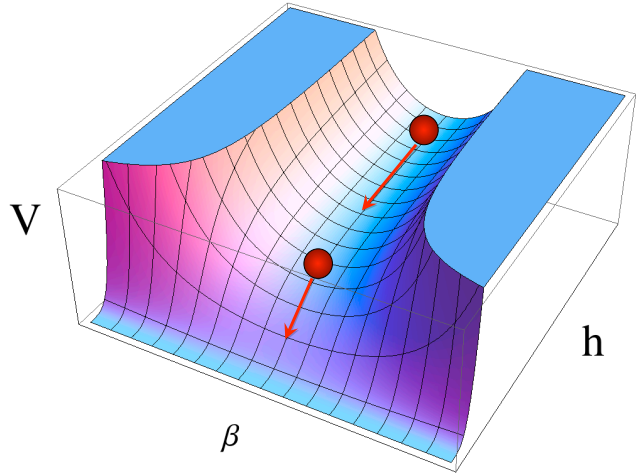}
\caption{During inflation at large $h$ the angular variable $
\protect\beta $ is stabilized at $\protect\beta =\protect\pi /4$,
corresponding to $h_{1}=h_2$. For $g^2,g^{\prime 2}\gg \lambda
^2$, this stabilization is preserved even after the end of
inflation.} \label{fig3}
\end{figure}

However, the angle $\beta$ may play an interesting dynamical role {\it
at the end of inflation}. Our calculations show that the potential
vanishes at $h = 0$ for all $\beta$, see Fig. \ref{fig3}. However, in
our investigation we did not take into account spontaneous symmetry
breaking in the SM, as well as soft terms leading to supersymmetry
breaking, which are important at an energy scale much smaller than the
energy scale relevant for inflation. Clearly, the low-energy scale
dynamics of the field $\beta $ will depend on the above mentioned
effects that we ignored, but also on the value of the field $\beta $ at
the end of inflation.

One could expect that until the low-energy effects become important, the
field $\beta $ remains equal to $\pi /4$. However, this is not always
the case. Indeed, (\ref{Sste}) yields that when $\chi h^2$ becomes
smaller than $O(1)$ and inflation ends, the mass squared of the field
$\beta $ becomes
\begin{equation}
m_\beta^2\approx h^2\left( -\lambda
^2+{\frac{g^2+g^{\prime }{}^2}2}\right) \,,
\end{equation}
thus affecting the slow-roll parameter $\eta _{_\beta }$ as follows
(recall Eq. (\ref{Ssste})):
\begin{equation}
\eta _{_\beta }\approx \chi \left[ \chi h^2\left( -1+{\frac{
g^2+g^{\prime }{}^2}{2\lambda^2}}\right) \right] \,.
\end{equation}
Note that  typically $|\eta_{_{\beta}}|\sim \chi \gg 1$, and that when
$\chi h^{2}$ becomes smaller than $O(1)$ and inflation ends. Therefore,
for $ g^2,g^{\prime }{}^2>2\lambda^2$, the $D$-term continues to
dominate the dynamics of the field $\beta $ even at the end of
inflation, $\eta _{_\beta }$ remains large and positive, and $\beta $
continues to be captured at its original value $\beta =\pi /4$, see Fig.
3. Oscillations of the inflaton field $h$ near the minimum of its
potential may lead to perturbative \cite{Dolgov:1982th,Abbott:1982hn},
as well as non-perturbative \cite{Kofman:1994rk,Kofman:1997yn} decay of
this field, which can be very efficient because the coupling constants
of the corresponding interactions are rather large. A detailed
discussion of reheating in the original (non-supersymmetric) version of
this scenario can be found in the second and third Refs. of
\cite{Sha-1}.

\begin{figure}[h]
\centering
\includegraphics[scale=0.35]{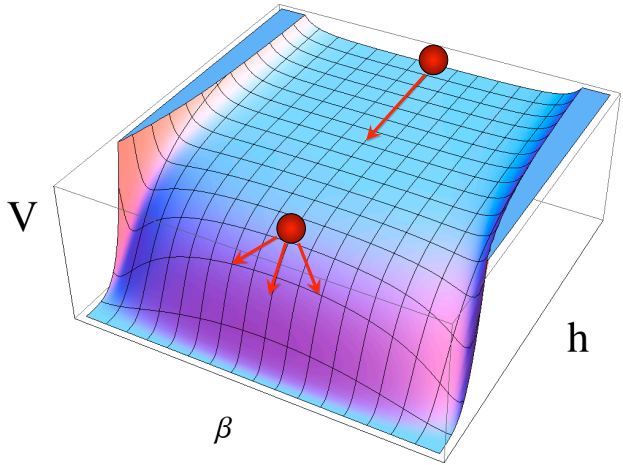}
\caption{During inflation at large $h$ the angular variable $\beta$ is stabilized at $\beta =\protect\pi /4$, corresponding to $
h_{1}=h_2$. For $g^2,g^{\prime 2}\ll \lambda^2$, at the end
of inflation the curvature of the potential in $\protect\beta
$-direction becomes large and negative, much greater than the
curvature in the inflaton direction. This leads to tachyonic
instability, generation of large fluctuations of the field
$\protect\beta $, and spontaneous symmetry breaking. } \label{fig4}
\end{figure}

The situation is more complicated in the opposite case $g^2,g^{\prime
}{}^2<2\lambda^2$, in which the field moving along the trajectory $
\beta =\pi /4$ experiences strong tachyonic instability at the end of
inflation, which leads to spontaneous symmetry breaking, see Fig.
\ref{fig4} . This effect, which is called ``tachyonic preheating''
\cite{Felder:2000hj,Felder:2001kt}, is similar to the waterfall regime
in the hybrid inflation scenario \cite{Linde:1991km}.

\begin{figure}[h]
\centering
\includegraphics[scale=0.35]{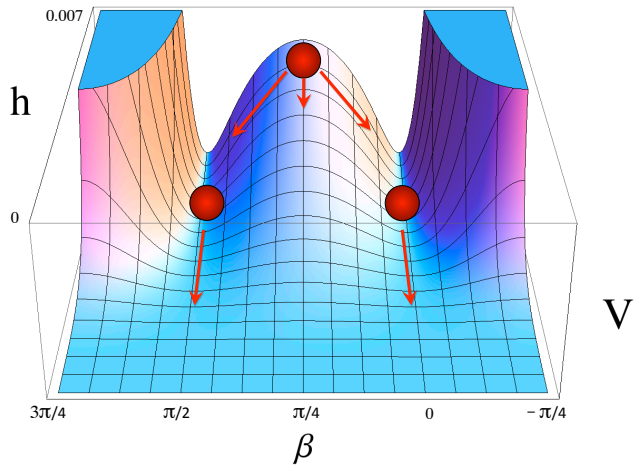}
\caption{``Tachyonic preheating'' effect at the end of inflation (for $
g^2,g^{\prime 2}\ll \lambda^2$).} \label{fig5}
\end{figure}

The physical meaning of ``tachyonic preheating'' within the framework
under consideration can be understood as follows. As mentioned,
inflationary regime ends when $\chi h^2$ becomes $O(1)$. At that time,
the parameter $ \eta _{h}$ describing the slow-roll in the $h$ direction
becomes $O(1)$, which means that the effective mass squared of the field
$h$ becomes $O(H^{{ -1}})$. Therefore the field $h$ reaches the minimum
of the potential at $h=0$ within the time $\Delta t=O(H^{{-1}})$ from
the end of inflation. This last, post-inflationary, part of the field
evolution is shown in Fig. \ref{fig5}.

During that time, quantum fluctuations of the field $\beta$ start
growing, $\delta\beta \sim \rme^{m_{\beta} t}$, they rapidly reach the
minima of the potential in the $\beta$ direction, which correspond to
the two valleys in Fig. \ref{fig5}, at $\beta \approx 0$ and at $\beta
\approx \pi/2$.  Spontaneous symmetry breaking occurs within the time
$m^{{-1}}_{\beta}$, which is shorter than $H^{{-1}}$ by the factor
$O(\eta^{{-1/2}}) \sim 10^{-2}$. In other words, this process occurs
almost instantly, on the cosmological time scale. When this happens, the
universe becomes divided into domains with the field $\beta$ taking
values in one of the two valleys in Fig.  \ref{fig5}. These domains, of
initial size $m^{-1}_{\beta}$, will be separated from each other by
domain walls corresponding to the ridge of the potential at $\beta =
\pi/4$. Then the field $h$ will continue rolling down to smaller values
of $h$, following the two valleys of the potential. A detailed evolution
of the field distribution can be studied by the methods developed in
\cite{Felder:2000hj,Felder:2001kt}.

In order to find out which of the two regimes ($g^2,g^{\prime
}{}^2>2\lambda^2$ \textit{versus} $g^2,g^{\prime }{}^2<2\lambda ^2 $)
occurs in the realistic versions of this scenario one should perform an
investigation of the running of the coupling constants from their
present day values to the end of inflation, similar to the investigation
performed in \cite{Sha-1}. However, prior to such an investigation, one
should find a solution to the main problem of this scenario, which is
the tachyonic instability with respect to the field $s$ found in Section
\ref{ss:sugraNMSSM}.

\end{document}